# CONVERGENCE
# &
# NEXT GENERATION NETWORKS

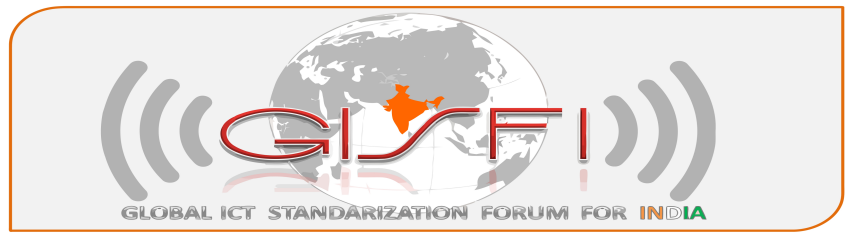

**WHITE PAPER**

**JULY 2009**

**Jaydip Sen**

**Tata Consultancy Services Ltd.**

**GLOBAL ICT STANDARDIZATION FORUM FOR INDIA**

**(GISFI)**

# Table of Contents

# List of Figures

# List of Tables

# Executive summary

The communications sector is undergoing significant changes, with the emergence of a number of platforms available to provide a different range of services. Some of these platforms are complementary to each other, while others are competitive, or can provide a valid substitute for some of the services provided. Up till now, the most important communications platform in most of the developing countries has been the public switched telecommunication network (PSTN) which provides access to all households and buildings. This universality in providing access has also meant that the network has generally been designated as one for universal service.

This chapter focuses on the area where the most significant changes are taking place in the communication sector. The objective of this paper is neither to give an overview of all communication platforms, nor is it aimed to assess the relative extent to which different platforms complement or compete with each other. The central theme of this paper is to examine the developments in what is commonly refereed to as next generation access networks and next generation core networks and their role in convergence. The focus on the next generation access and core networks is because of their huge impact and change these network will bring in telecommunication sector.

The chapter also discusses the issues of convergence of services like voice, video and data in the next generation networks and how it can lead to more competition in individual markets for each of these services. Convergence of services puts increasing competitive pressure on mobile carriers from the IP world. However, the trend towards horizontal integration of infrastructures, market and services could lead to strengthening of market power as there may be relatively few companies that can package voice, video and data services in a single bundled offer to end users.

In addition, the chapter presents the technical challenges for migration towards the converged next generation networks (NGN). This migration requires changes in the network topology which potentially involves several structural changes, such as a re-organization of core network nodes and changes in the number of network hierarchy levels. The shift to IP networks also raises questions of whether interconnection frameworks need to be revised, such as a shift to use interconnect frameworks which have been successful in developing Internet markets. With the range of technologies making demand on spectrum, such as HDTV, mobile TV or 3G services, growing rapidly, the need to have a framework for allocation and management that can flexibly reassign unused and underused spectrum is becoming increasingly becoming important. The paper briefly discusses this issue.

Several other issues related to interconnection and interworking in the converged NGN network such as numbering, naming and addressing issues, next-generation access technologies, lawful interception, fixed-mobile convergence, security and privacy issues of the users are also discussed. As an example of convergence in NGN, the paper discusses IP Multimedia Subsystem (IMS). The architecture of IMS and the functionalities of its various components are discussed. The activities of various standardization bodies in convergence and IMS are briefly discussed. A brief presentation is also made on the current and the emerging trends in the Indian telecommunication industry. Particular focus is paid on the challenges that will be faced by the industry to sustain its growth rate and deploy next generation networks and services. Finally, various challenges and open issues in convergence and network interoperability are presented

# 1   Introduction

Communication networks have become a key economic and social infrastructure in many counties in the world. The telecommunication network infrastructure is crucial to the national and international exchange of goods and services, and acts as a main catalyst in changing economic interrelationships through rapid technological change and the proliferation of a range of new services. With the development of the Internet, the role of communication networks has evolved and their importance increased. The advent of higher access speeds, in many cases symmetric speeds, available to business and to residential subscribers, has also increased the role of communication infrastructures by expanding the available range of services. High speed networks are increasingly helping resolve ongoing societal concerns in areas such as the environment, health care and education, and are increasingly playing a role in social networking. However, for the potential of new network technologies to be realized, the market will require that these networks have universal, or close to universal coverage. The full potential of networks is only likely to be achieved where markets are effectively competitive and solutions have adequate coverage to most geographic areas.

Technological innovation, stimulated through digitalization, has been a major factor in driving change in the communications market. This innovation is reducing costs and enhancing the capability of networks to support new services and applications. A key innovation which is expected to bring further significant changes in the communications market is the transformation from circuit-based public switched telecommunication networks to packet-based networks using the Internet Protocol, so-called *next generation networks* (NGN). NGN is expected to completely reshape the present structure of communication systems and access to the Internet. The present structure of vertically independent, although interconnected, networks may be transformed into a horizontal structure of networks based on Internet Protocol. Investment requirements for NGN are high and, as for any investment, there are risks. Policies need to ensure that risks and uncertain returns are compensated while ensuring competition since, without competition, the benefits of high speed broadband and NGN will not be realized.

The developments in new communication structures and the impetus they are expected to give to the present process of *convergence* in networks, services and terminals are expected to lead also to new policy challenges. Convergence, by changing service boundaries, service characteristics and stimulating the offer of new services, may require that new markets are regulated differently than existing ones. It remains to be seen to what extent the deployment of NGN and convergence will facilitate the process of creating durable competitive conditions in communication markets or will raise further obstacles to the creation of competition. It is fairly evident, however, that changes taking place as a result of investment in next generation access and core networks and the convergence of technologies, services and markets will require reviews and rethinking of existing policy and regulatory frameworks.

The rest of the chapter is organized as follows. Section 2 introduces the concept of convergence and NGN. Section 3 presents details of the access networks and core networks for the next generation converged applications and also describes the drivers of the NGN. Section 4 describes various possible interconnection frameworks for NGN deployment. Section 5 presents broadcasting convergence in all-IP networks. Section 6 discusses the security aspects of the NGN. Section 7 introduces IP multimedia subsystem (IMS) and presents its architectural details. Section 8 describes a case study based on IMS in residential networking environment. Section 9 discusses the details of various standard bodies involved in the activities related to convergence and NGN. Section 10 presents the current and future trends of the Indian telecommunication industry with particular attention on the challenges the industry will face in deployment of NGN and converged services. Section 11 presents some open research problems in convergence and finally Section 12 concludes the chapter.

# 2 Convergence and NGN

Convergence in network technologies, services and in terminal equipment is at the basis of change in innovative offers and new business models in the communications sector (Box 1). The utilization of the term "convergence" represents the shift from the traditional "vertical silos" architecture, *i.e.* a situation in which different services were provided through separate networks (mobile, fixed, CATV, IP), to a situation in which communication services will be accessed and used seamlessly across different networks and provided over multiple platforms, in an interactive way. Already in the 1990s, the possible impact of digitalization and convergence between telecommunications and broadcasting was under examination and proposals made for changes in existing regulation. The growing role of the Internet in the economy and society has enhanced the process of convergence and its rate of change.

---

**Box 1: What is Convergence?**

The path towards convergence was led mainly by the increasing digitalization of content, the shift towards IP-based networks, the diffusion of high-speed broadband access, and the availability of multi-media communication and computing devices. Convergence is taking place at different levels:

**Network convergence**- driven by the shift towards IP-based broadband networks. It includes fixed-mobile convergence and 'three-screen convergence' (mobile, TV and computer).

**Service convergence** – stemming from network convergence and innovative handsets, which allows the access to web-based applications, and the provision of traditional and new value-added services from a multiplicity of devices.

**Industry/market convergence** –brings together in the same field industries such as information technology, telecommunication, and media, formerly operating in separate markets.

**Legislative, institutional and regulatory convergence** – or at least co-operation- taking place between broadcasting and telecommunication regulation. Policy makers are considering converged regulation to address content or services independently from the networks over which they are provided (technology neutral regulation).

**Device convergence** – most devices include today a microprocessor, a screen, storage, input device and some kind of network connection-increasingly they provide multiple communication functions and applications.

**Converged user experience**- unique interface between end-users and telecommunications, new media, and computer technologies.

The process towards convergence has been based on an evolution of technologies and business models, rather than a revolution. This process has led to:

- Entry of new players into the market.
- Increasing competition among players operating in different markets.
- The necessity for traditional operators to co-operate with companies previously in other fields.

As a result, convergence touches not only the telecommunication sector, but involves a wider range of activities at different levels, including the manufacturer of terminal equipment, software developers, media content providers, ISPs, etc.

---

Previously distinct communication networks and services are today converging onto one network, thanks to the digitalization of content, the emergence of IP, and the adoption of high-speed broadband. Traditional services such as voice and video are increasingly delivered over IP networks and the development of new platforms is facilitating the provision of converged services (Table 1). These converged services are appearing in markets as "triple" or "quadruple" play offers which provide data, television, fixed and mobile voice services.

The process of convergence has also been facilitated by the opening up of telecommunication markets to competition. Although large telecommunication operators have played a role in the process of convergence, new market players have moved rapidly, and often in an unpredictable way, adopting different market models from traditional telecommunication firms. Voice over IP is a clear example of such services, disrupting traditional markets, pushing towards adoption of next generation networks and facilitating convergence. Internet service providers started offering VoIP as a cheaper way to communicate over the Internet. Services were offered on a "best-effort" basis by third parties, over any Internet connection. Today the market for VoIP services is varied, with network access operators providing VoIP as a replacement for PSTN voice telephony, often guaranteeing access to emergency services, or a certain QoS. Internet service providers continue to offer access to VoIP services from multiple platforms and from anywhere in the world. Mobile VoIP is also emerging, both as a service provided by the network operator or as an application that can be downloaded on any Wi-Fi enabled handset. Initiatives, such as Google's '*Android*', are likely to put pressure on existing mobile operators to charge flat rates for mobile Internet access, thus eventually increasing the degree of substitutability between mobile and fixed Internet access (in terms of price rather than speed) [1].

**Table 1. An all IP-based converged environment**

| Telecommunication environment | Next-generation converged environment |
|---|---|
| Single purpose networks | Multi-purpose networks |
| PSTN, cellular, broadcast | IP network (providing voice, video and mobile services) |
| Narrow-band | Broad-band |
| Vertical silos | Destroys compartmentalization <br><br> Traditional boundaries between industry segments (e.g., telephony, cable TV, broadcasting, wireless) are blurring- need to re-think market definitions (product definition and geographic boundaries definition) |
| Network-service link | New services and content developed independently of the network |
| Operators control services to end users | Increased consumer control |

On the content side increasing competition is taking place between network access operators, including wireless, cable or satellites, all offering video, music, or other content to their users. A growing number of operators are also focusing on mobile content, in particular on the possibility to download music, or access applications and online services from a mobile device. The possibility to provide video content over IP is often seen as a new way to propose content to users, and as an opportunity for network operators to enlarge the range of services they offer to their customers. Content services, especially those over managed IP networks, have still not exploited their full potential. In most cases, access to content is offered in a form very similar to traditional broadcasting, with defined timetables, geographical distribution, rigid copyright schemes, a very low degree of interactivity, and a traditional billing scheme although a number of operators are now beginning to offer more flexible programming with video on demand and distribution of video content from popular Internet sites. Changes are often taking place as a result of an increasing number of users creating and exchanging their own content on a multiplicity of devices, which imply a shift away from simple passive consumption of broadcasting and other mass distribution models towards more active choosing, interacting, the creation of content, and the emergence of a participatory culture [2]. These developments also increase the need to communicate, and the demand for symmetric communications.

With the growing offers from access platforms, and the different types of video services and applications available – digital terrestrial, IPTV, HDTV, Video on Demand, but also disruptive applications such as Joost [3] or Sling box [4] – the concept of "social value" of terrestrial broadcasting may become weaker, while the impact of these new services remains to be assessed.

Although the extent and effects of such convergence are yet to be seen, the phenomenon is already challenging the existing remit of many sector–specific domestic regulations. For example, the impact of convergence on competition is likely to be mixed. On the positive side, the move towards next generation networks, able to deliver a wide range of communication services, creates a schism in many traditional market definitions. While in the past telecommunication companies only offered fixed-line voice, and policy makers could easily define the specific market and make regulatory decisions, today the convergence of video, voice and data on next generation broadband networks can lead to more competition in individual markets for each of these services. As a result, convergence touches the telecommunication, cable television and broadcasting sectors, and involves a wider range of activities at different levels, going from manufacturers of terminal equipment, software developers, media content providers, ISPs, etc.

> **Box 2. Bundled services**
>
> Bundling refers to the sale of a number of services combined in a single price package, usually excluding the possibility that customers can obtain a single service without taking or paying for the other services in the bundle.
>
> Bundling of services can help generate economies for the supplier through, for example, reduction in service marketing charges, customer acquisition costs, billing charges etc. For the client bundling often has the advantages in that prices are lower compared to having to subscribe to each service individually, however, customers may not want all the services offered in a bundle. A client who does not want IPTV may be obliged to pay for these services when subscribing to certain triple play bundles.
>
> At the same time a bundle, while normally offering a better price than purchasing the same services separately, is also difficult to assess when trying to compare prices across a range of different offers. A service provider may also use a service in the bundle to cross-subside other services using this to obtain an unfair market advantage. Bundling may also make it difficult for regulators to define markets, assess market power, and therefore understand whether or not dominance exists in a given market.

On the other hand, the trend towards horizontal integration of markets and services could lead to strengthening of market power, as there may be relatively few companies in a country that can provide a combined video, voice and data offering. This may lead to a reduction in competition for the communications sector as a whole. In addition, bundling of services may make it more difficult to determine the extent to which prices are cost-oriented, allowing cross-subsidization between services. Service convergence and the shift towards next generation networks could therefore contribute to the creation of additional bottlenecks and control points, which may need to be addressed by the regulator (Box 2).

In this context, next generation networks (NGN) provide the technical underpinning of convergence, representing a single transport platform on which the carriage of previously distinct service types (video, voice, and data) "converges", together with new and emerging services and applications. While different services converge at the level of digital transmission, the separation of distinct network layers (transport, control, service and applications functions – Figure 1) provides support for competition and innovation at each horizontal level in the NGN structure. At the same time NGNs also create strong commercial incentives for network operators to bundle, and therefore increase vertical and horizontal integration, leveraging their market power across these layers. This may bring about the need for closer regulatory and policy monitoring, in order to prevent the restriction of potential development of competition and innovation in a next generation environment, and therefore the risk of reducing benefits for consumers and the potential of new networks for economic growth.

# 3   Next-Generation Networks

Although there is a significant amount of work underway in standardization forums on NGN, at the policy level, there is a still not complete agreement on a specific definition of "NGNs". The term is generally used to depict the shift to higher network speeds using broadband, the migration from the PSTN to an IP-network, and a greater integration of services on a single network, and often is representative of a vision and a market concept. From a more technical point of view, NGN is defined by the International Telecommunication Union (ITU) as a "*packet based network able to provide services including telecommunication services and able to make use of multiple broadband, QoS-enabled transport technologies and in which service related functions are independent from underlying transport-related technologies.*" NGN offers access by users to different service providers, and supports "*generalized mobility which will allow consistent and ubiquitous provision of services to users*" [5]. NGN, also defined as "broadband managed IP networks", includes next generation "*core*" networks, which evolve towards a converged IP infrastructure capable of carrying a multitude of services, such as voice, video and data services, and next generation "*access*" networks, *i.e.* the development of high-speed local loop networks that will guarantee the delivery of innovative services.

## 3.1 Next-generation access networks

The definition of next generation *access* networks is usually specific to investment in fiber in the local loop, *i.e.* fiber replacing copper local loops, able to deliver next generation access services – *i.e.* an array of innovative services, including those requiring high bandwidth (voice, high-speed data, TV and video). However, while next generation access networks tend to refer to a specific technological deployment, there are other technologies which can compete in providing some of the services which it is envisaged will be provided by NGNs. There are also other technologies which may not be able to fully compete with NGN access networks in terms of capacity and the plethora of bundled offers which NGNs can provide, but may be perfectly suitable for users who do not have the need for high capacity access. The different technologies available include existing copper networks upgraded to DSL, coaxial cable networks, powerline communications, high speed wireless networks, or hybrid deployments of these technologies. Although fiber, in particular point-to-point fiber development, is often described as the most "future proof" of network technologies to deliver next generation access,[6] there are likely to be a number of alternative and complementary options for deployment of access infrastructures by incumbent telecommunications operators, and new entrants.

**Cable:** Cable television (CATV) operators have begun to upgrade their infrastructure to hybrid fiber copper (HFC) allowing for bidirectional traffic and using Docsis [7] technology to increase network capacity. These developments are allowing CATV companies to offer voice and Internet access (data services) in competition with telecommunication companies which through their offer of Internet TV have begun to compete with CATV companies. Offering data and voice services, in addition to television, helps cable companies differentiate their product offering from satellite providers. The bandwidth provided by cable networks, using Docsis 3.0, will allow for 160 Mbit/s downstream and 120 Mbit/s upstream for end-users.

**Broadband wireless access:** Broadband wireless access (BWA) technologies aim at providing high speed wireless access over a wide area. Certain early fixed wireless access technologies, such as *local multipoint distribution service* (LMDS) and *multi-channel multipoint distribution service* (MMDS), never gained widespread market adoption. WiMAX technologies, – the IEEE 802.16 set of standards that are the foundation of WiMAX certification, and similar wireless broadband technologies, are expected to address some of these shortcomings, and fill market gaps left by wired networks, or compete with wired access providers. The WiMAX Forum has estimated that new WiMAX equipment will be capable of sending high-speed data over long distances (a theoretical 40 Mbit/s over 3 to 10 kilometers, in a line-of-sight fixed environment). When multiple users are connected, capacity sharing will significantly reduce speeds for individual users sharing the same resource [8].

Wi-Fi (or wireless fidelity) refers to wireless local area networks that use one of several standards in the 802.11 family. Wi-Fi allows LANs to be deployed without cabling for client devices, typically reducing the costs of network deployment and expansion. Due to its affordability, scalability and versatility, its popularity has spread to rural and urban area. Wi-Fi range is usually limited to about 45 m indoor and 90 m outdoors, however Wi-Fi technologies can also be configured into point-to-point and point-to-multipoint networks in order to improve their range and provide last mile fixed wireless broadband access. One way to serve remote areas which cannot be reached with the above-mentioned technologies is with wireless "mesh" solutions. They often include a satellite backhaul connection through Very Small Aperture Terminals (VSAT), usually coupled with wireless technologies such as Wi-Fi. This combination allows access to telecommunication and data services even to more remote areas, albeit with limited (and expensive) bandwidth [9].

Terrestrial wireless services offer the opportunity to deploy competing access infrastructure. However, they may offer different service characteristics to fixed-line services in terms of coverage, symmetry and speeds. These networks may be less suitable to deliver sustained high bandwidth connections for larger numbers of users, or for high bandwidth applications, such as High Definition TV on demand. In addition, wireless service deployments are constrained by spectrum availability. At the same time, the economics of their deployment is often relatively scalable, which means that they have lower economic barriers to entry compared to fiber deployments [10]. While they may not be a complete substitute, they can complement wireline networks for specific services [11].

**Broadband over powerlines (BPL):** Use of the power grid as a communications network, or "power-line communications" appears to provide a series of advantages, offering not only voice, but also broadband services, with the connection speed not dependent on distance from the telephone exchange (as happens with DSL) or number of customers connected (as with cable). With this system a computer (or any other device) would need only to plug a BPL "modem" into any outlet in an equipped building to have high-speed Internet access. Notwithstanding the benefits that the availability of an extensive infrastructure can allow, for the moment the service provision is far from standardized [12], and the capacity of bandwidth provided through BPL is still being questioned.

**3G mobile networks:** The term NGN frequently encompasses some kind of fixed-mobile convergence (FMC) [13], as it allows the transition from separate network infrastructures into a unified network for electronic communications based on IP, which facilitates affordable multiple business models, seamlessly integrating voice, data and video. The introduction of 3G technology supports the transmission of high-speed data with speeds theoretically reaching 2/4 Mbit/s. The operators are expanding their 3G networks across different countries. This will provide higher data speeds to users enabling them to access innovative networks dedicated to provide mobile video or television programming. However, existing 3G technologies will need to be upgraded in order to support very high bandwidth or extensive concurrent usage that may be demanded by users in the future. The future evolution of mobile networks for example using LTE technology (Long Term Evolution) – a next generation mobile technology – may significantly increase speeds, enabling high peak data rates of 100Mbit/s downlink and 50Mbit/s uplink.

**Satellite networks:** Satellite services are typically dedicated to direct-to-home television and video services, satellite radio, and specialized mobile telephony uses. More recent technological advances such as spot beam technology and data compression algorithms have enhanced the technical efficiency in satellite communications, enabling more efficient use of spectrum, and reduced redundancy. It has resulted in increased effective data density and reduced required transmission bandwidth [14]. Satellite broadband is usually provided to the customer via geosynchronous satellite. Ground-based infrastructure includes remote equipment consisting of a small antenna and an indoor unit. Gateways connect the satellite network to the terrestrial network. Except for gateway locations, satellite broadband is independent of terrestrial infrastructure such as conduits and towers, allowing it to provide coverage also to remote areas.

## 3.2 Next-generation core networks

The next generation *core* networks are defined on the basis of their underlying technological "components" that include – as mentioned in the ITU definition – packet-based networks, with the service layer separated by the transport layer, which transforms them into a platform of converged infrastructure for a range of previously distinct networks and related services. These features may have an impact on traditional business models and market structure, as well as on regulation:

- *IP-based networks*: "Next generation core networks" generally cover the migration from multiple legacy core networks to IP-based networks for the provision of all services. This means that all information is transmitted via packets. Packets can take different routes to the same destination, and therefore do not require the establishment of an end-to-end dedicated path as is the case for PSTN-based communications.

- *Packet-based, multi-purpose*: While traditionally separate networks are used to provide voice, data and video applications, each requiring separate access devices, with NGN different kinds of applications can be transformed into packets, labeled accordingly and delivered simultaneously over a number of different transport technologies, allowing a shift from single-purpose networks (one network, one service), to multi-purpose networks (one network, many services). Interworking between the NGN and existing networks such as PSTN, ISDN, cable, and mobile networks can be provided by means of media gateways.

- *Separation of transport and service layers*: This constitutes the key common factor between NGN and convergence, bringing about the radical change in relationship between network "layers" (transport infrastructure, transport services and control, content services and applications). In next generation networks service-related functions are independent from underlying transport-related technologies (Figure 1). The uncoupling of applications and networks allow applications to be defined directly at the service level and provided seamlessly over different platforms, allowing for market entry by multiple service providers on a non-discriminatory basis.

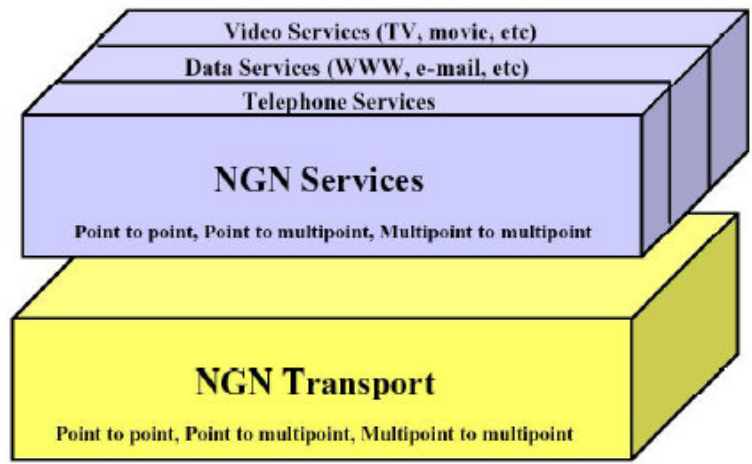

**Figure 1. Separation of functional planes**
(Source: Keith Knighston, Industry Canada, ITU NGN Architecture, presentation at the "ITU-IETF Workshop on NGN', May 2005)

These features may foster the development and provision of new services and constitute a new opportunity for innovation, allowing different market players to create value at the separate functional levels of access, transport, control and services.

However, while initially it was a common assumption that this layered structure would lead to a market model where services could be increasingly provided across the value chain, in a more decentralized manner, today it appears that the network provider will decide whether the "horizontal" model will prevail, or whether they will simply (commercially) vertically integrate transport and services across functional levels, offering bundled services [15].

Currently, bundling of a variety of services is a key trend in the sector, bringing greater competition between formerly distinct sectors. Bundles include all sorts of combinations of fixed and mobile voice calls, Internet access and media/entertainment services (Section 5). With services and transport commercially integrated at the vertical level, customers are somehow "locked-in" in a vertical relationship with a single operator. This is not negative in itself, as packages are often more convenient, or easier to use, at the same time it is important to maintain the possibility for users to choose which services they want to purchase, and to have clear information about the cost and characteristics of these. The risk would be to create a situation in which the network provider may limit the possibility of users to access IP-based services and applications provided by third parties.

Considering the economic drivers behind the shift towards next generation networks, there is an incentive for the network provider to also become an integrated market player, in order to maintain/extend their user base or benefit from a privileged relationship with subscribers. This raises questions regarding obligations for access to networks by service providers and issues of traffic prioritization [16]. In this context access plays an important role for all service providers to be able to provide their content, services and applications to end users.

One essential feature of next generation networks is the capability to support "*generalized mobility which will allow consistent and ubiquitous provision of services to users*" [17]. Although core next generation networks tend to be on a fixed infrastructure, the possibility to improve interconnection with mobile networks is being explored, and standardization organizations as well as operator and manufacturers associations are working to the development of appropriate standards. In addition, the deployment of wireless infrastructures facilitates access to IP networks, and the adoption of increasingly sophisticated devices and handsets will allow an easy access to IP services from anywhere.

The migration process towards IP-NGN potentially entails several structural changes in the core network topology, such as the rearrangement of core network nodes and changes in the number of network hierarchy levels. As a result, an overall reduction in the number of points of interconnection will take place, especially with regard to interconnection points at the lowest level. This could negatively affect alternative operators whose previous interconnection investment may become stranded [18]. For example, BT today has some 1 200 exchanges at which competitors have installed DSLAM's, using local loop unbundling to provide broadband and bundled services. In addition, BT has over 700 exchanges at which competitors can connect their voice services. The number and location of points at which competitors could connect their networks to BT's voice services is expected to reduce substantially to at most 108 Metro-node sites, and probably to a subset of these which could number as few as 29, while the number and location of exchanges at which local loop unbundling is likely to be possible are not expected to be affected by the roll-out of 21CN [19].

## 3.3 NGN drivers and impact

NGN is an evolutionary process and it can be expected that operators will take different migratory paths, switching to NGN while gradually phasing out existing circuit networks, or building a fully-IP enabled network from the outset [20]. The investment in developing NGN is motivated by several factors (Table 2). Telecommunication operators across most of the countries in the world have been faced with a decline in the number of fixed-line telephone subscribers, coupled with a decrease in *average revenue per user* (ARPU), as a result of competition from mobile and broadband services [21]. Traditional sources of revenue (voice communications) have declined rapidly and fixed-lines operators are subject to an increase in competitive pressure in the market to lower tariffs and offer innovative services. This has generated pressure from the investors' community to decrease the cost and complexity of managing multiple legacy networks, by disinvesting from non-core assets and reducing operational and capital expenses.

In this context, the migration from separate network infrastructures to next generation *core* networks is a logical evolution, allowing operators to open up the development of new offers of innovative content and interactive, integrated services, with the objective to retain the user base, attract new users, and increase ARPU. NGN is therefore often considered essential for network operators to be "more than bit pipes"[22] and to strategically position themselves to compete in the increasingly converged world of services and content, where voice is no longer the main source of revenue, and may become a simple commodity. The investment in next generation access networks – both wired and wireless – will be necessary in order to support the new services enabled by the IP-based environment, and to provide increased quality. At the same time, the important investment necessary to develop next generation infrastructures brings about new economic and regulatory issues, which will be analyzed in the following sections.

**Table 2. NGN drivers [23]**

| Economic Drivers | Technological Drivers | Social Drivers |
|---|---|---|
| • Erosion of fixed line voice call revenues<br>• Competitive pressure from new entrants in high-margin sectors of the market (long-distance, internal) and vertically integrated operators (triple-play bundles).<br>• Saturation of both fixed and mobile telephone services<br>• Retain and expand users' base, lower customer churn<br>• Ability to expand into new market segments<br>• Possibility of 'ladder of investment', i.e. a phased approach for investment, initially targeting more densely populated areas, and then gradually | • Obsolescence of legacy networks, plus cost and complexity of managing multiple legacy networks<br>• Lower capital and operational expenses. Increased centralization of routing, switching and transmission, lower transmission costs over optical networks.<br>• IP-based networks enable the provision of cheaper VoIP services as a replacement for PSTN voice services.<br>• IP-based networks enable the provision of a wider range of services, and allow bundling of services (triple and quadruple play).<br>• Evolution and convergence of terminal | • Demand for innovative, high-bandwidth, services (HDTV, VoIP, etc).<br>• Demand for more targeted or personalized content (on demand multimedia services, mobility),<br>• Demand for increased interactivity: possibility to interact actively with the service, growing interest for user-created content.<br>• Demand for evolved and more flexible forms of communications, including instant messaging, video-conferencing, P2P, etc.<br>• Business demand for integrated services, in particular, in case of multi-national structures, which need to link different national branches, guaranteeing a flexible and secure access to centralized resources and |

Although the shift in the migration to all-IP networks is taking place at different paces in different countries, several operators have already updated their transport networks, and are now dealing with NGN at the local access level. Solutions embraced by fixed operators may also increasingly support IP Multimedia Subsystem (IMS), to enable fixed-mobile convergence [24].

For the moment the most common services provided through the new networks are the provision of PSTN/ISDN emulation services, *i.e.* the provision of PSTN/ISDN service capabilities and interfaces using adaptation to an IP infrastructure, and video on demand (VoDs). At the same time the business world is showing an increasing interest in new NGN-enabled services and applications. Companies are migrating their Time Division Multiplexing switches to IP in order to enable integrated applications for specific industry-based functionalities and purposes [25].

Progress in the field of mobile (cellular) communications is taking shape with the development of the IMS standard [26]. For the moment two services have been standardized under the IMS protocol, Push to Talk over Cellular (PoC) and Video Sharing [27]. Prominent telecommunication network equipment suppliers are actively supporting the take up of IMS and some of them are implementing IMS strategies and commercial IMS products [28]. IMS is seen as the enabler for the migration to next generation networks of mobile operators and therefore for the implementation of fixed-mobile convergence. No evident killer application has currently emerged, with many operators focusing on one specific service: voice. Facilitating the use of voice applications, enabling users to handle their calls easily between fixed and mobile networks, and to receive calls wherever they are, is fundamental for the take–up of the service. Operating in an IMS environment would allow a seamless handover from WLAN (fixed) to mobile during calls (Voice Call Continuity).

In order for real-time voice calls to be offered seamlessly between the circuit switched domain and the Wireless LAN interworking with IMS architecture, the Third Generation Partnership Project (3GPP) [29] is currently working to develop the appropriate Technical Specifications to define this functionality as a standard 3GPP feature. The study by 3GPP of the standard is underway [30]. In the meanwhile, fixed-mobile converged services have been launched by some mobile operators with access to fixed networks, using a different standard – Unlicensed Mobile Access (UMA) [31] – allowing users to seamlessly switch from fixed to mobile networks (see below, paragraph on Fixed Mobile Convergence).

In addition, increasing competitive pressure on mobile carriers is coming from the IP world. Thanks to the availability of dual-use devices and Wi-Fi hotspots, service providers – such as Skype, Google, and others – are able to offer on the market a host of new services for mobile users in a very short period of time. This rapidity constitutes an important comparative advantage, which in some cases provoke the reaction of mobile operators (and manufacturers), tending to limit the services and applications users can access from their mobile handset.

### 3.4 Internet and NGN

Technological developments associated with next generation networks should help combine the characteristics of the traditional telecommunication model, and of the new Internet model, dissolving the current divisions and moving towards a harmonized and coherent approach across different platforms, gradually bringing to full convergence fixed and mobile networks, voice, data services, and broadcasting sectors. In short, in the future the choice of the technology used for the infrastructure or for access will no longer have an impact on the kinds and variety of services that are delivered.

This however does not reflect the current situation, where the two worlds still have different visions and commercial models (Figure 2).

The telecommunications tradition emphasizes the benefits of higher capacity local fiber access facilities, and powerful network intelligence. Access in this context should be simple and reliable, with centralized network management and control to guarantee the seamless provision of a wide range of services, bundled network-content-applications offers, and one-stop shop solutions.

On the other hand, the Internet world traditionally focuses on edge innovation and control over network use, user empowerment, freedom to choose and create applications and content, open and unfettered access to networks, content, services and applications. Freedom at the edges is considered more important than superior speed of managed next generation access networks.

Indeed, the "Internet" still represents different things to different people, and next generation networks are seen as both a possibility for improved services or as a way to constrain the Internet into telecommunication boundaries, adding new control layers, capable of discriminating between different content, and "monetize" every single service accessed.

Services provided over next generation networks will differ from services currently provided over the public Internet which is based on a "best effort" approach, where the quality of transmission may vary depending on traffic loading and congestion in the network, while with NGN packet delivery is enhanced with Multi Protocol Label Switching (MPLS). This allows operators to ensure a certain degree of Quality of Service – similar to the more constant quality of circuit switched networks – through traffic prioritization, resource reservation, and other network-based control techniques, as well as to optimize network billing as in circuit-switched transport [32].

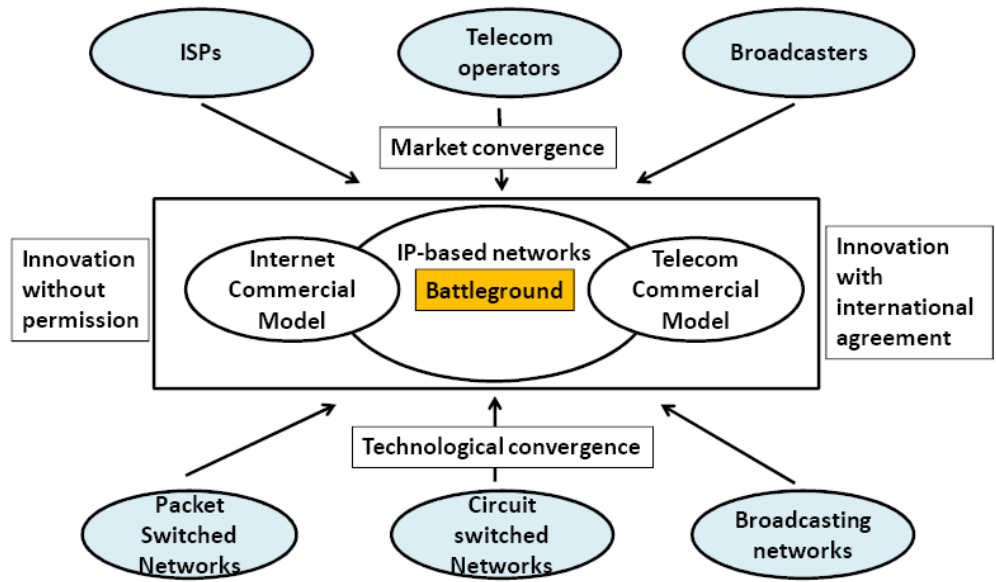

**Figure 2. The convergence model**

(Source: J. Horrocks, "NGN and Convergence Models, Myths, and Muddle", OECD NGN Foresight Forum, 3 October 2006)

The concept of network-based control seems to be the main difference between the public Internet approach and next generation managed IP networks approach. NGN offers the possibility to provide a detailed service control and security from within the network, so that networks are aware of both the services that they are carrying and the users for whom they are carrying them, and are able to respond in different ways to this information. In contrast, the Internet aims to provide basic transmission, remaining unaware of the packets/services supported. While the Internet model remains therefore completely open to users and new applications and services, in managed IP networks operators are able to control the content going through the network [33]. In turn, this may have negative implications for the content of third party providers if their traffic is discriminated against in relation to that of an integrated operator.

## 4  Interconnection Frameworks

Interconnection is essential in a competitive communications environment since it provides the means to allow the customer of any one communication service provider to connect with the customer of any other communications provider, and any service provider to connect, and provide service, to a customer irrespective of their network carrier. The transition to IP-based next generation networks is likely to raise questions as to how interconnection should be take place, given the significant differences in interconnection practices between the PSTN and IP networks and the fact that there will be interconnection between diverse networks including cable networks and the development of new services such as fixed-mobile converged services.

In the PSTN environment traditionally service providers adhere to wholesale payment arrangements known as *calling party's network pays* (CPNP), where the network of the party that places (originates) a phone call makes a wholesale payment to the network of the party that receives (terminates) the call. In contrast, Internet interconnection has been based on *peering*, *paid peering*, and *IP-transit*. With peering, two *Internet service providers* (ISPs) agree to exchange traffic solely among their respective customers, sometimes without payment; with transit, one ISP agrees to carry the traffic of a customer (possibly also an ISP) to third parties, generally for a fee. These arrangements based on commercial agreements result in an interconnected Internet, and have generally not been subject to regulatory obligations. The model that applies is therefore determined, in practice, by the type of interface used to exchange the traffic. The question is therefore on which model interconnection in a converged NGN environment should be based.

Depending on the strategy of companies, there will be a transition phase during which it is likely that both sets of practices will coexist as the proportion of IP traffic increases, and that of circuit-switched traffic decreases. This may imply, as well, a transition in interconnection procedures. In many countries regulators use *long–run incremental cost* (LRIC) models to determine interconnection costs. There is a need for regulators to assess how the two sets of interconnection arrangements operate in their current milieus to evaluate whether these should be maintained in an NGN environment. The market for exchange of IP traffic, as regards the Internet, has worked well, producing efficient arrangements and lower prices, and allowing for entities of different sizes to exchange traffic [34].

In terms of physical facilities supporting traditional fixed and mobile switched interconnection, the migration towards NGN changes the network topology which potentially involves several structural changes, such as a re–organization of core network nodes and changes in the number of network hierarchy levels [35]. As an example in Germany Deutsche Telekom has 74 nodes for its IO network compared to 475 nodes for the PSTN [36]. This may lead to a geographical re–arrangement of points of interconnection, and to the reduction in the number of points, especially at the local level. At the same time it can result in new entrants being subject to stranded investment requiring them to invest in new infrastructure in order to reach new points of interconnection. There are different fiber network topologies in a NGN access environment which also may need to be taken into account since the requirements and points of interconnection may differ [37].

The separation of networks functional planes should allow for the creation of a horizontal platform for the provision of services, separated from the transport layer. For this separation to be effective, interconnection should be possible at all functional levels. However, there is the risk that operators do not consider horizontal separation appropriate, as it is more difficult to guarantee a certain level of quality of service in interconnected networks, or simply because it is not in their best interest. Most incumbent operators still see NGN as a simple continuation of vertically integrated transport and services, as in the case of legacy networks [38].

## 4.1 Numbering, Naming and Addressing

Telephone numbers, domain names, IP addresses, and other addresses are crucial resources for communication and access to the market. They provide operators and service providers with the necessary data for locating and identifying customers and network points in order to deliver their services. For end users they provide a presence in the world of communication and a means to communicate with others. For the PSTN, the public switched telephone network, the telephone numbering system [39], is the core mechanism to address end users. Practically all wire line and wireless networks operators base their interconnection, interoperability and service provisioning on the telephone system. With NGN, the existing numbering system is expected to continue, at least in the short to medium term, as the dominant scheme within voice communication to identify and connect subscribers.

Nevertheless, the same developments that characterize the merging communications landscape, such as the migration to IP, are affecting addressing as well, which raises risks in that access for users to competing service providers and/or services of their choice might not be achieved if the resolution between both addressing systems used (telephone numbers in PSTN, and IP addresses, domain names and *uniform resource identifiers* (URIs) in Internet) is not properly addressed with global standardization [40].

The IPv4 addressing scheme [41] as used in the Internet has been universally embraced by NGN networks as the core new addressing scheme, in combination with the overarching TCP/IP protocol suite [42]. IP addresses are used 'under the hood' within networks and determinate unique network points; using an IP address will always lead to the exact location of that network point. On top of IP addressing there are translation mechanisms, such as the DNS (Domain Name System) that map or add other identifiers to an IP address. These identifiers, such as domain names, e-mail addresses and SIP addresses [43], are more comparable to telephone numbers, as they are used at the edges of networks, in the higher layer where services and applications take place in interaction with users.

With the expansion of the public Internet, the use of domain names and e-mail addresses for end users has become common practice worldwide, comparable to the expansion and acceptance of the telephone numbering system. Increasingly the underlying general format used in IP networks is the URI, the Uniform Resource Identifier. The URI is evolving into the main intra-network identifier and basically defines an 'identity-service' combination in a format like scheme:user@host or scheme:identifier@domain.tld. The URI format is versatile and, next to the well known URI for e–mail (mail to:user@domain.tld), the URI for SIP (sip:user@host) is becoming a main identifier to address VoIP subscribers according to the SIP protocol. These types of identifiers are all IP-based and can eventually be traced back to an IP address.

In parallel, other more closed identifier schemes have been introduced, mainly with the emergence of web–based VoIP and instant messaging (IM). Internet-focused companies such as eBay (Skype), Microsoft, Yahoo, Google and AOL have added voice, IM (instant messaging) and video capabilities to their software, serving large communities. They route mostly on the basis of 'end to end point' communication, having the advantage that traffic does not need to be routed through the PSTN's traditional switches, or via SIP gateways as used within VoIP. These highly competitive providers on the voice market manage their subscribers' identities with proprietary schemes [44] and employ telephone numbering only when interoperability is needed with subscribers outside their community (Skype-in).

Although implemented on a provider by provider basis, IP–based schemes follow a standardized format and can be in principle supported across other networks. Interoperability is feasible if there is agreement between providers. The absence of interoperability is sometimes seen as a deliberate customer 'lock in', as concluded by some parties on the basis that, *e.g.* Skype, will not map their end users to URIs, and the introduction of IP telephones that cannot be used for anything other than the application provided by the IP telephony provider.

Telephone numbers by which PSTN subscribers are identified may eventually evolve into alternative names and addresses, but generally many new services, such as web–based IM and VoIP services, are used 'on top' of the regular voice subscription and this does not lead to the substitution of telephone numbers The emergence of new addresses, however, does lead to increasing divergence, as users are collecting more numbers and identifiers in different schemes, but there are no real indications that this divergence is posing problems on the end–user side; end–user equipment is becoming more intelligent and capable of handling multiple addresses and managing contact details.

The divergence however, does pose a challenge for providers. Telephone numbers in their standard format are not supported in the core NGN networks based on IP, where generally the URI format or other IP-based identifiers are used. Still, for users as well as for providers, being able to continue to use telephone numbers is considered crucial for the shift from the classic telephone service to VoIP and for the integration of new IP multimedia services. ENUM [45], a standard developed by the IETF [46] was conceived for this purpose; it offers a mechanism for transforming public telephone numbers into unique domain names. While solving the mapping problem, it introduced potential new applications, as a result of the insertion in the Domain Name System.

ENUM comprises a set of standards and mechanisms for transforming public telephone numbers into unique domain names to be used in NGN, enabling providers and users to continue to use telephone numbers which is considered crucial for the shift from the existing public switched telecommunication environment to an Internet Protocol based environment and is thus becoming an essential building block for NGN embedded. Due to ENUM the lifespan of the existing telephone numbering scheme could be prolonged, subsequently maintaining the role of telephone numbers as key identifiers for telecommunication services. Eventually, however, regulators may need to introduce more flexibility in numbering plans by broadening the uses for existing number ranges, and considering portability of numbers between different services. At the same time access to ENUM data will become crucial to set up interconnection.

## 4.2 Universal Access and Next-Generation Access

Convergence and the transition to next generation networks could, in the longer term, have an impact on the definition and scope of *universal service obligations* (USOs). At present USOs focus on the provision of voice services [47]. USOs generally refer to the requirement that a designated USO telecommunications operator provides a minimum set of services (which include voice telephone service) to all users, regardless of their geographical location within the national territory, at an affordable price, even though there may be significant differences in the cost of supply. Differently, the term "universal access" is used to refer to a situation where every person has a reasonable means of access to publicly available network facilities and services.

The communications market has been subject to significant changes both in terms of the means to provide voice services (mobile, VoIP) and the decreasing importance of voice services as a proportion of total telecommunications usage (*e.g.* because of e-mail, SMS, etc.). Many countries have stressed the economic and social importance of broadband access which in turn has led to considerations as to whether broadband access should be included as part of USOs. As the communications market evolves, particularly with regard to next generation networks, policy makers may need to review definitions of universal service to determine whether changes need to be made and, if so, what services and access would be required, and whether funding mechanisms should change.

The goal of universal service obligations generally are to promote the "availability, affordability and accessibility" [48] to telecommunications services. Definitions of universal service across most countries are relatively similar although there are differences in the mechanisms used to achieve these goals. Implicit in universal service goals in many countries is national tariff averaging aimed at assisting rural households (on the assumption that service costs are higher in those areas). In many countries part of USOs include, as well, special tariffs for those on low incomes.

Internet access is, to some extent, already included in universal service. For example, in the United States, the federal universal service schools and libraries program provides, among other things, discounts for Internet access for schools and libraries throughout the nation, while the federal universal service rural health care program provides, among other things, discounts to ensure comparability in Internet access rates paid by health care providers in rural areas and urban areas. In addition, the Federal Communications Commission (FCC) has initiated a universal service rural health care pilot program, which seeks to stimulate deployment of the broadband infrastructure necessary to support innovative tele-health and, in particular, telemedicine services to those areas of the United States where the need for those benefits is most acute. The European *universal service directive* (USD) [49] specifies that connections to the public telephone network at a fixed location should be capable of supporting speech, fax, and data communications at rates sufficient for "functional Internet access." The provision of functional Internet access has been interpreted by the Directive as encompassing simply the provision of a "narrowband connection", [50] and no minimum data rate is mandated in the directive. Overall, it seems that most EU countries opted for not requiring more than a 28Kbit/s connection.

The definition of universal service is an evolving concept which may change over the years, to reflect advances in technologies and usages. For example, in the United States, universal service specifically is defined as "an evolving level of telecommunications service that the [FCC] shall establish periodically . . .taking into account advances in telecommunications and information technologies and services" [51]. In the EU, to ensure that the changes in USO designations justify the important associated policy interventions, the Universal Service Directive established a number of criteria for modification. These usually include the popularity of the service, the diffusion of the technologies, and the likeliness that the unavailability of the service causes social exclusion. They also include considerations regarding "technological feasibility", the possibility to find "practical and efficient implementation mechanisms", and the balance between the cost of the measure and the benefits it will brings to society, always seeking to minimize market distortions [52].

## 4.3 NGN Lawful Interception

Subject to national legislation, all kinds of telecommunications may be subject to interception and/or data searches in relation to enquiries. *lawful interception* (LI), also called "wiretapping", consists in the interception of communications by *law enforcement agencies* (LEAs) and intelligence services. The requests are directed to public telecommunication networks and services, in accordance with national legislation and on the basis of the authorization from competent authorities. With technological evolution it has become more difficult to intercept all communications of a targeted user. In the *public switched telephone network* (PSTN) environment interception was carried out by connecting to the line of the user at the local switch. With the advent of mobile phones, it became more difficult to implement lawful intercept since users could be at any location served by the operator and its roaming partners. The mobile signaling networks need to be monitored to detect the presence, identity and location of callers. On the technical side standards organizations have played a role in formulating standards which allow for lawful interception.

Convergence of networks and services, with users transmitting information through IP-based, mobile or fixed networks interchangeably, is exacerbating the problem of lawful intercept. To preserve the ability of law enforcement agencies to conduct electronic interception, network operators and application service providers, as well as manufacturers of telecommunications equipment, are required to modify and design their equipment, facilities, and services to ensure that they have the necessary capabilities to intercept. Governments extended the obligation to provide lawful interception from network operators to include also Internet service providers [53]. However, considering that often the reference is not anymore the connection, but the service used over the connection, questions arise as to whether the coverage of lawful intercept is adequate and whether this requires retention of data by, for example, Internet Service Providers. Communications using instant messaging or e-mail, as an example, do not necessarily have to be 'home-based', but can use web-based mail where servers are located outside a country so that cross-border enforcement also becomes important. In this context, it is essential for law enforcement authorities to co-operate with network and service providers, [54] as well as with application service providers, and to continue to work at the international level to build effective co-operation networks among different countries.

## 4.4 Fixed-Mobile Convergence

In the future network technology such as IMS (IP Multimedia Subsystem), should provide a standardized next generation architecture based on Internet Protocol (IP) for operators, and allow for the provision of mobile and fixed services using converged handsets embedding a radio interface such as cellular/Wi-Fi or cellular/Bluetooth dual-mode handsets. Currently, the main factor promoting FMC is the trend towards VoIP-enabled wireless telephony (VoWi-Fi), *i.e.* devices that use Wi-Fi to connect to a VoIP service such as Skype or roam between cellular and wireless LAN systems. Some of the VoWi-Fi operators are at present providing Wi-Fi based only services, but some are starting to offer FMC services by combining cellular services with VoWi-Fi. Challenges to mobile telecommunications operators are also coming from Wi-Fi hotspot operators, such as Boingo, allied with Skype. Some mobile operators are linking or considering linking their cellular networks with Wi-Fi hotspots and using VoWi-Fi to improve indoor coverage and offer low-cost calling in Wi-Fi locations. At present there various ways being used to provide FMC services, some of which are more technologically integrated than others. Dual-mode cellular/Wi-Fi handsets and using Wi-Fi modems in the home environment to access VoIP through ADSL connections can be found in some countries. There are less evolved forms of FMC using cellular/Wi-Fi dual-mode handsets that do not have a handover function or have a handover function but do not utilize a fixed voice or broadband network in the home. Services also exist linking both fixed and mobile networks which are not technologically converged, such as those offering a single voice mailbox over both fixed and mobile networks.

The deployment of NGN is expected to accelerate the offer of FMC services which are seamless to the user and use least cost routing. In turn, this may require that regulators review existing frameworks to ensure that they are not a disincentive to the development of new services, and that existing frameworks treat new services in a technologically neutral way. Numbering policies also have to accommodate FMC services and, if existing geographic numbers are used, then, in a calling party pays system, it may be necessary to devise ways to inform the call originator if different charges will be assessed based on the called party's location. It may also be important for regulators to develop adequate market tests given that the incumbents already have market power and often their mobile operators are also the market leaders; the development of FMC can augment this market power.

# 5   Broadcasting Convergence into IP-Based Networks

The digitalization of content, added to the shift towards IP-based networks, the diffusion of high-speed broadband access, and the availability of multi-media devices, allowed an increasing convergence of broadcasting and telecommunication sectors. The production and diffusion of audio-visual content does not seem to be limited to traditional broadcasters anymore. Telecommunication operators are providing content along with Internet access, newly emerging providers are offering access to content over IP, and traditional broadcasters are crossing over to other platforms, transmitting their programs also over IP networks.

Furthermore, the development of next generation mobile services-using 3G and 4G networks, or mobile broadcasting systems – enables the delivery of high quality audiovisual (AV) content to portable devices and mobile phones. Convergence is nowadays a reality, with different types of content and communication services delivered through the same pipes and consumed over a variety of platforms and user devices. Convergence over multiple access platforms has not only affected the distribution market, but also created new forms of usage, providing consumers with greater choice and control over content. Multimedia, interactive audiovisual services are increasingly transforming users from passive watchers of TV programs to active players able to decide what they want to see, when and on which device. Video on Demand, Personal Video Recorder (PVR) services, peer to peer (P2P), or user–created video, therefore, herald an important change in the traditional broadcast model to exchange audiovisual content among large audiences. Media consumption, tastes and preferences may become more fragmented, the importance of social networks as a means to participate in content creation will probably continue to grow, and there will be an increasing demand for new types of content, able to fully capture the new capacity of the Internet for interactivity, non-linear consumption and participation [55].

The evolution of technology does not necessarily change many of the social and cultural broadcasting policy objectives, but technology may change the way that they are presently implemented and may allow for increased market liberalization than that which has been common in the sector while allowing the core policies to be maintained. The digitalization of transmission, for example, enables a more efficient use of spectrum than analogue transmission, increasing significantly the number of terrestrial broadcasting channels which can be made available. When analog TV signals are switched off, a significant amount of spectrum bandwidth will be freed up, and will be available potentially for other applications, such as mobile television, high-definition television, mobile broadband networks and WiMAX networks [56]. Audio-visual content providers may include network operators, which are usually provide digital television and content over IP networks as part of their "triple play" bundles, or new service providers, such as Joost [57], using P2P technologies to stream content over the Internet, or YouTube, based *inter alia* on user created content. Broadcasters are also entering the IP market, launching new content platforms, such as Hulu-a NBC/NewsCorp venture (Table 3).

**Table 3. Examples of different ways to access content in a converged environment**

| | Provider | Content | Business model | Upload of user created content | Geographic restrictions |
|---|---|---|---|---|---|
| **Managed IP Networks** | France Telecom | DTT + Vo D (DSL) | Commercial + subscription channels | No | Yes |
| | BT Vision | DTT + VoD (DSL) | Commercial + subscription channels | No | Yes |
| **New Internet Service/ Application Providers- IPTV Model** | Joost | Streaming Independent/ Private content producers | Ad-supported | No | Yes- geographical blocking depending on content rights |
| | Babelgum | Streaming Independent/ Private content producers | Ad-supported | No | Yes- geographical blocking depending on content rights |
| **Broadcast Operators** | Hulu (Beta) NBC/New Corp | Premium content from NBC/Fox + content from 15 other cable channels. Streaming from the main site or distribution partners (AOL, Yahoo, Comcast, MSN, MySpace) | Ad-supported, banners alongside the video, text along the bottom of the picture or clip | No | Yes: cannot access the service from outside the US |
| | BBC | On demand 7 day catch-up of BBC TV and radio programming | - | No | - |
| | YouTube | User created content, short professional video/trailers, promotional materials | Ad-supported targeted adverts, banners, etc. | Yes | No |
| | iTunes | Download of movies, music and podcasts, distribution agreements with content producers | Pay per download, free content is also available | Through podcasts | Yes, cannot download movies outside the US |

As the market for audiovisual services becomes more dynamic, content producers will be able to offer services directly to all new markets without intermediaries or gatekeepers. With content available on new platforms and networks, there should be lower entry barriers, and the sector could become more open and competitive over the next years. At the same time this will bring up the issue of the need for network neutral policy approaches, for both fixed and mobile networks, in order to avoid the creation of barriers to access for independent service providers [58].

In addition, existing government instruments to control broadcasting content – such as quotas for protection of language and culture, pluralism requirements, or must carry obligations-are challenged by the new multiplatform environment, and may need to be adjusted in order to continue to fulfill their goals.

Convergence not only leads to a larger and more competitive market, but also a more international market. A globally structured market – in terms of ownership, investment, and distribution and marketing strategies – offers an enormous potential to the media industries, but also poses new challenges to national regulation, which may not always be compatible across borders, therefore risking to be less effective, not enforceable, or – if excessively restrictive – to slow down growth of media players in an international content market.

## 5.1 Convergence in Content

While convergence may contribute to plurality and diversity, as it lowers market entry barriers, it creates new issues and challenges to existing policy. The telecommunication and broadcasting policy traditions may need to adjust in order to cope with the changing markets and to continue to achieve common policy objectives.

- **Scope of regulation:** Audiovisual content is increasingly distributed via a broad range of digital technologies that transmit to television, computers, as well as mobile and portable devices, blurring boundaries between "video" and "broadcasting services" [59]. The scope of the definition of broadcasting services[102] is relevant considering the detailed regulation which is usually imposed on broadcasters and usually aimed at addressing a number of social and economic interests, such as the need to maintain plurality and cultural diversity, develop national identity, and implement certain standards of decency. Policy makers need to determine whether and to what extent existing broadcast regulations should apply or be adapted to a wider range of content packagers and suppliers, and to what extent existing broadcast regulation may be reduced.

- **Ensuring effective competition:** Convergence is helping to intensify competition in broadcast markets by impacting on delivery networks and services. Convergence can help reduce access bottlenecks by allowing services to be delivered on a number of different platforms, and by creating market entry opportunities by new providers stimulates innovative services. The entry into the audio-visual market by new players, such as telecommunication network operators and larger Internet-based companies, can reduce market power in broadcasting. However, access to content is important for new entrants so that if larger companies or joint ventures (horizontal integration) control media rights for the most interesting premium content, it may be difficult for new entrants to provide competitive offers [60]. In addition, the development of some of the new technologies and services depends on the spectrum which is made available. With the shift to digital television more spectrum will be freed up and will be available for other services. The allocation of the so–called "digital dividend" can therefore have an impact on the development of new services in the content market [61]. Currently it seems that the request for spectrum will be driven by mobile television (Box 3), high-definition television and wireless services, such as WiMAX.

- **Spectrum allocation:** The switch off of analog TV signals and the shift to digital transmission will make a significant amount of spectrum bandwidth available (the so-called "digital dividend"), which could be used for the provision of enhanced TV services, more TV channels, or some advanced wireless communication services [62]. In particular, the availability of spectrum to develop new wireless networks could help new entrants to create alternative access infrastructures and deliver directly their services to users, competing with incumbent operators.

> **Box 3. Mobile video content**
>
> The limitations and the cost of offering television on 3G networks using multimedia broadcast multicast service (MBMS) have encouraged operators to try to obtain separate allocations of spectrum for mobile television using a number of technologies. In addition, the interest for the allocation of new spectrum bandwidth may also be a means of pre-empting competition from broadcasters offering mobile television, and push convergence from the network into the handset.
>
> The mobile television market does not seem to have deployed its full potential yet and innovation has been lagging behind, with sometimes restrictive platforms adopted by the wireless carriers and phone manufacturers. The EC estimated that the market for mobile TV would reach EURO 20 billion by 2011. However, it seems that mobile operators still have difficulties in identifying the appropriate business model for the service. Currently, revenues for mobile TV mainly come from the subscriptions, as advertising is not expected to be significant because of the low usage.
>
> In November 2007, Google announced the launch of a new mobile operating system called Android. Based on Linux, Android provides an open platform for developers to create their own applications for a wide range of mobile devices, and will be available for free to cell phone manufacturers. Mobile-tailored content with targeted advertisement could therefore be one of the future models for mobile television. Another model for mobile television could be Qualcomm's one-way, multicast video programming network, MediaFLO. The MediaFLO service is currently offered by one mobile operator in the United States in approximately 40 US markets.

- **Public interest objectives for content:** The rationale for special regulation of broadcast content is changing along with digitization and increasing access to on-demand audio-visual services. There is more choice, and an increasing proportion of consumers can now control the time of consumption of content. It is important, in view of the changes in the supply of information and programming, to reconsider how public interest objectives can be achieved in the digitalized IP world.
- **Advertising:** Advertising quotas and time frames were developed to limit commercial communications in traditional linear, point to multipoint broadcast transmissions. In a more interactive environment, and with VOD and PVR providing some possibility to skip frames, the traditional advertising model has become less effective. A controlled liberalization of some rules for television advertising, such as product placement, interactive online selling and banners during certain programs, could help the development of new business models, allowing broadcasters to compete with innovative Internet-based video services.
- **Must carry regulations:** Some countries enforce certain form of *must-carry* regulation. Most of these regulations were formed when there scarcity in distribution networks. As a result of technological and market developments, there is less dependence on a single infrastructure, and more channels and platforms for distribution of content are now available. Must carry rules should therefore be limited to a reasonable number of channels, including especially public service channels. Instead of "must carry", consideration could be given to a framework whereby terrestrial broadcast channels should be subject to "must offer" requirements, *i.e.* certain broadcasters are obliged to offer their content to other platforms if they ask for it.

## 5.2 Emerging Trends

Customers are now driving the market. It is the changing life style of the users that is making the device vendors to come up with devices that gives the flexibility. The new devices (iPhone, Palm Pre,

Win Mobile) are giving the choices to the customers to drive the market. iPhone, for example, is a disruptive technology which created its own single eco system making the users build applications on it. The requirement of having one device for any kind of activity like voice calls, SMS, MMS, browsing, gaming, contents while being on any network and requiring a single bill for the usage is now taken fro granted. This is making the services providers to move to the next generation systems, which is the convergence of devices, networks, and businesses.

Next generation operational support systems (OSS)/ business support systems (BSS) are being evolved to manage the converged IP networks. The service providers' mission as a surviving service provider in 2009 and beyond is to leverage assets to bring customer winning products to market faster and better than their competitors at a price point that will keep customers "sticky". Doing so means providing and maintaining ubiquitous connectivity and service delivery to a burgeoning line-up of consumer and business devices-laptops, TVs, mobile handsets, PDAs, MP3 / personal entertainment units- each with an array of functions that grows with introduction of each new model. The providers must enable users to mix, match and bundle services- whether that includes voice, video, data, wireless, entertainment, hosting and messaging or premium /lifestyle contents- at home, at work or in transit, and be ready to receive orders however customers wish to place them.

Next-generation *operations support systems* (OSSs) and *billing support systems* (BSSs) hold the key to service providers' new Holy Grail, which is the ability to deliver "any service to any device over any network, anytime, anywhere." But it takes a *service delivery platform* (SDP) to open the door.

Figure 3 presents an integrated OSS/BSS system. It consists of various components such as: InAdaptors, OutAdaptors, Transformers, Web Services, queues and BPM tool. A brief description of the components is provided below:

- **Business process management (BPM) layer:** it is used to define automated business processes. BPM systems can interact with the external systems through its off-the-shelf or custom-developed connectors. There are various players offering BPM tool like IBM, BEA, Vitria and Jboss. Most of the BPM tools deploy the business processes on a J2EE application server as *enterprise java beans* (EJB) and the whole application is assembled into an EAR. So these businesses processes have access to all the J2EE server features like security transaction, JNDI, remote connectivity etc.

- **Application queue:** the business processes defined in BPM can be invoked by sending a message in the application queue. The application queue is deployed as JMS queue on application server in order to provide asynchronous communication.

- **Connector:** it consists of InAdapter, OutAdpter, Web Service, Transformer and a connection-specific queue. Connector can be invoked either by placing the request in queue or by directly calling the web service. Connector is the only way through which the BPM layer can interact with the external system.

- **Connector Queue:** all the asynchronous requests that need to be handled by OutAdpter are placed in the Connector queue. Each connector has its own queue that is deployed as IMSqueue on the application server.

- **InAdaptor:** it is used to place a request in application queue that will invoke a workflow of BPM depending on the request type. For example, if account information has been updated in CRM system by CSR, it should be synchronized with the system. In order to accomplish the CRM INAdaptor will place an *updateaccount* request in the application server.

- **OutAdaptor:** it is used to process the requests that are placed in connector queue. In the CRM example, InAdaptor will place an *updateaccount* request in the application queue. Based on the message type, this request will go to the Billing queue, and then OutAdaptor will read the request, and call the web service to create an account in the billing system.

- **Transformer:** it is used to convert the object of a third party system into application-specific object and vice versa. It has only the transformation logic and no business logic. It should be able handle the following transformation: (i) Java object to Java object, (ii) Java to XML or vice versa, and (iii) XML to XML.
- **Web service interface:** the functionality of the external systems can be invoked through the webservice interface. The Web Service interface will accept the application-specific parameters and transform them to external system-specific parameters and will then call the functionality of the external system. Similarly, the result will be transformed and returned.
- **Customer/Self care portal:** it consists of custom GUI that interacts with Integration Module.
- **VNO:** a virtual network operator (VNO) who wants to use the services provided by another network access provider (NAP) may directly interact with either exposed web service layer or the BPM layer.

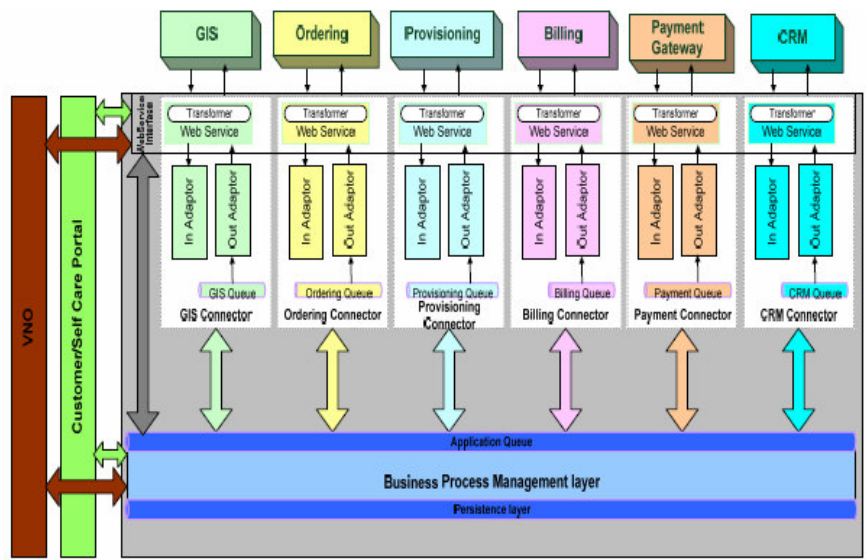

**Figure 3. Integrated OSS/BSS architecture**

# 6   Security in Converged NGN

The convergence of networks towards all-IP architecture provides operators with great opportunities to reduce their costs, and develop integrated services across fixed and mobile access increasing subscriber welfare. Network convergence needs to be complemented by convergence in the underlying security of policies, measures and practices so as to defend against various attacks [63]. As operators move from trials to wide commercial roll-out, questions regarding how to guarantee security across multiple networks are becoming more urgent.

IP-based next generation networks and the traditional circuit switched networks operate in different environments and are therefore exposed to different types of threats and attacks, both from within or outside the networks. With converged networks, operators are migrating from a stand-alone *closed* environment, such as the PSTN, to an open environment. The PSTN infrastructure is controlled by operators, and users have a lesser amount of information on its structure and functioning, as well as fewer possibilities to misuse the network. This situation, sometimes labeled as *security by obscurity*, [64] stands in contrast with the design of the IP infrastructure, based on open protocols, which were not originally designed for security implementation [65]. IP networks enable the provision of services

– such as voice, data, and multimedia-provided by multiple access and service providers, and are connected with a growing number of devices.

Security in a converging environment is not only a technical, but also an economic and social issue. On the economic side, networks are an integral part of the global information infrastructure, defined as an essential, indispensable facility for society, whose disruption would rapidly bring about a state of emergency or could have adverse societal effects in the longer term [66]. While governments and businesses are eager to adopt innovative services and applications; they require appropriate levels of assurance to protect their information and transactions. The social dimension of security is also important as convergence of networks and systems will expand opportunities for consumers to be connected anytime, anywhere. While the growing empowerment of users enables them to benefit more from ICTs, it also brings along with it increasing security and vulnerability risks for their transactions and personal information.

The borderless nature of IP networks means that security threats affecting the converged infrastructure can arise from anywhere. The main challenges across borders include the necessity to improve cooperation of law enforcement activities against security offences, with particular attention to consistency of cyber-crime legislation and regulations. In addition, international co-ordination and exchange of information is essential to create a global understanding of security risks and solutions linked to converged networks.

Although security is a priority in the future networks, it is also important to ensure an appropriate balance between civil liberties and security solutions – at the technical, policy or regulatory levels - in order to avoid excesses leading to violation of users' privacy, or illegitimately limiting individuals' rights to anonymity and freedom of expression [67]. It is also important to take into account the direct and indirect costs which may be incurred from securing networks. These costs also reduce the openness of networks and may impact on innovation.

International *standard development organizations* (SDOs) such as ITU, ETSI, ISO, IETF, 3GPP/3GPP2, are currently working to integrate security into the definition of NGN standards and protocols, in order to appropriately address security in the design phase of the new generation of networks. A set of specifications for IMS standards has been included in IMS Release 7, while TISPAN, in the preparation of its NGN Release 1, has been working on an equivalent set of specifications for broadband fixed access. TISPAN aligned its security approach with 3GPP where convergence was identified, adding TISPAN-tailored security specifications in areas where there are differences between fixed and mobile architecture. For example, pure wireline solutions do not have the same vulnerability as the mobile interface, which allows for the introduction of simplified security scenarios; on the other hand, fixed networks have to support inter-working with many sets of more or less secure protocol stacks, and with a wider variety of access technologies compared to mobile operators. In addition, user equipment vulnerability is more pronounced in fixed than in mobile networks, as users can modify their equipment without prior notice to the provider.

In a layered architecture, such as that of NGN, where services are separated from transport and access is enabled from multiple devices, security has to be considered at different points in the NGN architecture. In its NGN Release1, ITU stressed the need to provide security of end-users communications across multiple-network administrative domains and identified three security layers: infrastructure security, service security and application security [68].

NGN solutions vendors also address the problem of security at different layers. These include access security, addressing direct or indirect connectivity of networks to user equipment (UE); intra-domain security, which is under the responsibility of the operator of the domain in question; and inter-domain security, i.e., security risks and threats associated with interconnection with untrusted networks [69]. In the latter case, security policies [70] from the originating network are usually enforced towards the destination network domain thanks to the utilization of *security gateways* (SEG) situated at the borders of different domains and communicating during interconnection.

**Table 4. Threats and risks in VoIP**

| Threat | Risk Issues |
|---|---|
| Eavesdropping through interception an/or duplication | Access can be gained through any access point to the voice network (particularly if there are wireless access points in the same network that supports the VoIP service). Once access has been gained, network sniffer tools are commonly available to intercept IP-based traffic. |
| Loss, alteration or deletion of content | Exposure to programmed attack e.g., programmed substitution of Dual-Tone Multi-Frequency (DTMF) or Interactive Voice Response (IVR) |
| Caller ID/location may not be identified in an emergency | Complex numbering schemes, combined with incorrect PSTN access point routing, may provide wrong location information to emergency services. There is a greater risk of this happening when calls from remote offices are routed over a wide area network (WAN) before reaching the PSTN. |
| Lack of capacity/system management | Other network traffic can impact on VoIP traffic. |
| Denial of service attack | Swamping of network traffic resulting in no capacity to support voice. Can be targeted from within the enterprise or externally. |
| Viruses and other malware | Swamping of network traffic resulting in no capacity to support voice. Can be targeted from within the enterprise or externally. Viruses can also target specific VoIP protocols. |
| Power failure | VoIP is different from traditional telephony in that voice services are potentially vulnerable to a number of power failure points within the data network, e.g. local router and switches. In contrast, traditional telephony handsets are powered from one centralized point, usually with a backup battery bank. |

Source: Trusted Information Sharing Network (TISN) "Security of Voice over Internet Protocol: Advice for Chief Information Officers", September 2005. URL: http://www.dcita.gov.au/communication_for_business/security/critical_infrastructure_security.html.

**Example:** A specific example of possible security issues in an NGN environment can be provided by Voice over IP services. Voice is a critical service which in the past has benefited from separate PSTN and mobile networks, and had a certain degree of reliability. Shifting from PSTN to IP, the existing redundancy may be lost due to network convergence, and VoIP may inherit many of the problems already experienced by TCP/IP protocol data communications, such as attacks on confidentiality, integrity, availability and authenticity. Some of the current threats include transmission of viruses and malware, eavesdropping, denial of service (DoS) attacks (Table 4). Although, operators are currently working on secure solutions for VoIP, service providers believe that it may be difficult to implement security while maintaining an appropriate level of Quality of Service (QoS), because of the extra processing and possible delay in communication it may cause [71].

An issue which may need to be specifically addressed in the context of NGN security is *identity management*, which in the NGN field has been technically described (at the working level) as the management by NGN providers of trusted attributes of an entity such as a subscriber, a device or a provider [72]. In a converged environment users would be able to use a single authentication mechanism (sign-in) on any access point on the NGN. The development and implementation of an authentication mechanism that allows a single and secure identification while protecting the privacy of the users however, is a great challenge [73]. In an environment with multiple providers, a common authentication process is difficult to achieve. However, a common authentication mechanism is also crucial in order to maintain a relationship between users, devices, and service and access providers. In addition, interoperable identity management is an issue that spans all layers from infrastructure to applications, and requires both technical and regulatory approaches harmonized at the international level [74].

# 7 IP Multimedia Subsystems (IMS)

Recently, web-based multimedia services have gained popularity and have proven themselves to be viable means of communications. This has inspired the telecommunication service providers and network operators to reinvent themselves to try and provide value added IP-centric services. There was need for a system which would allow new services to be introduced rapidly with reduced capital expense (CAPEX) and operational expense (OPEX) through increased efficiency in network utilization. Various organizations and standardization agencies have been working together to establish such a system. Internet Protocol Multimedia Subsystem (IMS) is a result of these efforts. IMS is an application level system. It is being developed by 3GPP (3$^{rd}$ Generation Partnership Project) and 3GPP2 in collaboration with IETF (internet Engineering Task Force), ITU-T (International Telecommunication Union-Telecommunication Standardization Sector), and ETSI (European Telecommunications Standard Institute) etc. Initially, the main aim of IMS was to bring together the internet and the cellular world, but it has extended to include traditional wireline telecommunication systems as well. It utilizes existing Internet protocols such as SIP (Session Initiation protocol), AAA (Authentication, Authorization and Accounting protocol), and COPS (Common Open Policy Service) etc, and modifies them to meet the stringent requirements of reliable, real-time communication systems. The advantages of IMS include easy service quality management, mobility management, service control and integration.

At present a lot of attention is being paid to providing bundled up services in the home environment. Service providers have been successful in providing traditional telephony, high speed Internet and cable services in a single package. But there is very little integration among these services. IMS can provide a way to integrate them as well as extend the possibility of various other services to be added to allow increased automation in the home environment.

This section extends the concept of IMS to provide convergence and facilitate inter-working of the various bundled services available in the home environment, which may include but is not limited to communications (wired and wireless), entertainment, security etc. In this section, a converged home environment is presented which has a number of elements providing a variety of communication and entertainment services. The proposed network would allow effective inter-working of these elements, based on IMS architecture. The objective is to depict the possible advantages of using IMS to provide convergence, automation and integration at the residential level.

## 7.1 IMS Architecture

The IMS comprises a core network (CN) which is a collection of signalling and bearer related network elements. These CN elements operate collectively to provide multimedia services to the end user. The IP multimedia services are based on the IETF defined standards for session control and bearer control. The IMS terminal connects to CN via an IP-Connectivity Access Network (IP-CAN), which functions merely as a means to transport IP data. This allows IMS to achieve *access independence* as defined in 3GPP (TS 22.228 V7.2.0 [75]. The access independence refers to the ability for the subscribers to access their IP multimedia services over any access network capable of providing IP-connectivity, e.g. via:

- 3GPP (UTRAN, GERAN)
- Non-3GPP access with specified interworking (e.g. WLAN with 3GPP interworking)
- Other non-3GPP accesses that are not within the current scope of 3GPP (e.g. xDSL, PSTN, satellite, WLAN without 3GPP interworking)

To understand the functionalities of IMS the following definitions will be useful [75]:

- **IP Multimedia CN subsystem:** comprises all CN elements for the provision of IP multimedia applications over IP multimedia sessions.

- **IP Multimedia application:** an application that handles one or more media simultaneously such as speech, audio, video and data (e.g. chat, text, shared whiteboard) in a synchronized way from the user's point of view. A multimedia application may involve multiple parties, multiple connections, and addition pr deletion of resources within a single IP multimedia session. A user may invoke concurrent IP multimedia applications in an IP session.

- **IP Multimedia service:** an IP multimedia service is the user experience provided by one or more IP multimedia applications.

- **IP Multimedia session:** an IP multimedia session is a set of multimedia senders and receivers and the data streams flowing from senders to receivers. IP multimedia sessions are supported by the IP multimedia CN subsystem and are enabled by IP connectivity bearers (e.g. GPRS as a bearer). A user may invoke concurrent IP multimedia sessions.

The high-level requirements of IMS are as follows [76]:

- **Support to establish IP Multimedia sessions:** IMS can provide the users with a variety of services but the most basic and most important service is the audio and video communication. The IP multimedia session is designed to support one or more multimedia applications. The IMS system also ensures that there is no compromise or reduction in privacy, security, or authentication as compared with traditional systems.

- **Support for QoS negotiation and assurance:** The IMS provides support for QoS negotiation for IP multimedia sessions, both at the time of establishment and during the session by the user and the operator. It also ensures that end-to-end QoS for voice at least as good as that achieved by the circuit-switched wireless systems.

- **Support of interworking with the Internet and CS domain:** Support for interworking with the Internet domain is an essential requirement. The IMS users will be able to access information, services and applications available through the Internet. The IMS users will have the ability to establish IP multimedia session with non-IMS users from the Internet and the existing circuit-switched systems including PSTN and cellular networks.

- **Support for roaming:** IMS will allow users to roam between different service providers' networks. There are established procedures to transfer signalling, authentication, accounting, and other service-related information between different IMS operators in a standardized and secure fashion.

- **Support for service deliver control by the operator:** The system requires strict control in terms of service delivery options. The operator will have control over all the services being offered to the users. General control policies enable the operators to monitor and control the bandwidth requirements in the network. By defining individual policies, the operators will have the ability to control sessions based on authorization.

- **Support for non-standardized rapid service creation:** IMS services need not be standardized. This allows the service providers or application developers to economically and rapidly develop and deploy services which would work equally well in different networks.

### 7.1.1 IM CN Subsystem Architecture: Nodes and their functions

Before describing the IMS architecture, it is important to remember that IMS does not standardize a network element, but the functionality provided by the element. The manufacturer is free to decide about the physical design of the functional unit; two or more may be combined if deemed necessary. On the same lines, IMS does not standardize services but the service enablers.

The core network elements are shown in Figure 4. These nodes communicate with each other using specific protocols; each of these interfaces is identified using a reference point label as shown in Figure 4. A detailed list of all the interfaces and their operation is available in 3GPP TS 23.002 [77].

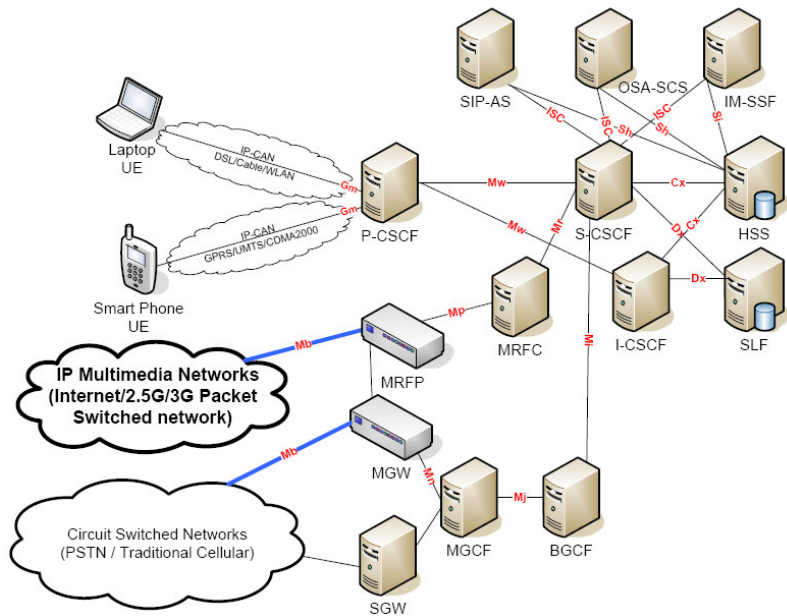

**Figure 4. IMS architecture general overview**

Another important characteristic of the IMS architecture is that it exclusively uses IPv6; it requires network elements such as NAT-PT (network Address Translation – Protocol Translation) and IMS-ALG (IMS Application Level Gateway) to interoperate with the traditional Internet (which mostly uses IPv4). The description of the nodes and their role in the IMS environment is briefly presented below.

#### 7.1.1.1 IMS Nodes: Databases

IMS utilizes one or more databases. The Home Subscriber Server (HSS) is used to store all the user related information, which is required to establish and handle multimedia sessions. The user information may include items such as the user profile (including the services the user subscribes to), location information, security information, the allotted S-CSCF address, etc. all the information is stored in a standard format and decisions are made by the HSS about the user sessions based on these items or user information. There could be more than one HSS, if the number of users is too high or for redundancy. In this case a Subscriber Location Function (SLF) is used to locate the HSS where the user information is stored. The SLF is a very simple data base which maps the user's address with an HSS, where all the user information is stored.

Both the HSS and SLF use DIAMETER protocol defined in RFC 3588 [78]. The DIAMETER protocol is the base protocol, there are specific applications developed for IMS to make the necessary decisions in the matters of authentication, authorization and accounting for a particular user.

#### 7.1.1.2 Serving Call Session Control Function (S-CSCF)

The S-CSCF is a SIP based server and is one of the three types of *call session control functions* (CSCF). The basic function of a CSCF is to process the SIP (Session Initiation Protocol) signaling in the IMS [79]. SIP provides all the functionality to establish and manage multimedia sessions over IP networks. The S-CSCF is considered as the central node in the signaling plane. It is basically a SIP server but performs session control as well. It also maintains a session state as required by the network operator. Within a network there could be a number of S-CSCFs, with different functionality and used for different purposes. The main functions of the S-CSCF as defined in 3GPP TS 23.228 [80] are mentioned below:

**Registration related operations**

- It may perform the task of the Registrar as defined in RFC 3261 [79]. It accepts the registration requests from the users, verifies the request by downloading the authentication vectors from the HSS. DIAMETER protocol is used for this purpose over the Cx interface.

- It downloads the user profile from the HSS, which contains the service profile or information about any application servers which need to be included in the SIP procedures.

- The S-CSCF makes the information available to the location servers, thus linking a particular user to an S-CSCF for the duration of the registration.

**Session related and session un-related flows**

- It controls the sessions for the registered users and might deny establishment of different sessions (IMS communication) on the basis of various conditions or clauses that bar such an activity for that particular user.

- The S-CSCF may behave as a proxy server as defined in RFC 3261[79] or subsequent versions of the protocol. It may accept and service requests locally or forward them to the relevant node after translation and filtering the request.

- The S-CSCF has the ability to behave as a User Agent as defined in RFC 3261 [79]; it can terminate and generate SIP transactions independently.

- It can interact with different service platforms or application server over the ISC (IP Multimedia Subsystem Service Control) interface. This interface allows for the coordination and support of various services provided by the application servers.

- The S-CSCF provides the endpoints with different service related information such as notification, location of additional media resources, billing notification etc.

- The functionality can be classified into two types; services provided for the originating end point and the for the destination end point. First we look at the services provided for the originating end point which may include an originating user/User Equipment or an Application Server (AS).

    o The S-CSCF obtains the address of the Interrogating CSCF (I-CSCF) of the destination user from the destination name. This is done in the case of the destination user being the customer of a different network. The S-CSCF forwards the request to the corresponding I-CSCF.

    o If the destination user belongs to the same network the S-CSCF forwards the SIP request or response to the allocated I-CSCF with in the network.

    o The S-CSCF also forwards the SIP request/response to a SIP server that is not a part of the IMS (e.g., internet). This depends on the policies of the home network operator.

- The S-CSCF forwards the SIP request or response to a Breakout Gateway Control Function (BGCF) for routing calls or sessions to the PSTN of any other Circuit Switched Domain (e.g. traditional cellular providers).

- If the incoming request is from an Application Server (AS), the S-CSCF will verify the request coming from the AS is an originating request and proceed accordingly. The S-CSCF will process and proceed even if the user on whose behalf the AS is acting is not registered and reflect in the charging information that AS initiated the session on behalf of the user.

- Services for the destination endpoint (terminating user/UE)

    - The S-CSCF will forward the SIP request or response to a specific Proxy CSCF (P-CSCF) as a part of the SIP terminating procedure to a home user within a home network or for a roaming user in a visited network.

    - The S-CSCF will forward a SIP request or response to an I-CSCF as part of the SIP terminating procedure for a roaming user in a visited network, where the home network operator chooses to include the I-CSCF in the path.

    - The S-CSCF will modify the SIP request as per directions for the HSS and the service control interactions, for routing the incoming session to the CS domain. This allows the user to receive the incoming session via the CS domain. It also forwards the SIP request or response to a BGCF for call routing to the PSTN or a CS domain.

    - The SIP request might contain preferences for the characteristics of the destination endpoints, the S-CSCF performs preference and capability matching as specified in RFC 3312 [81].

**Charging and resource utilization management monitoring**

- Like all the other nodes of IMS, the S-CSCF is a part of the complicated charging or accounting procedure. It generates Charging Data Records (CDR) for this purpose.

The S-CSCF is always located in the home network, and there are usually a number of S-CSCFs in a network for the sake of scalability and redundancy. Each of them can serve a number of IMS terminals at the same time.

### 7.1.1.3 Proxy-Call Session Control Function (P-CSCF)

As shown in Figure 4, the P-CSCF is the first contact between the user and the IMS network in the signaling plane. All the signaling and control information passes through the P-CSCF before getting to the user. It acts as an outbound/inbound SIP proxy server. The discovery of the address of the P-CSCF and its allotment to the user is performed during the process of IMS Registration and this does not change for the duration of the registration. The P-CSCF discovery process is described in section 5.1.1 of 3GPP TS 23.288 [80].

The various functions performed by the P-CSCF are mentioned below:

- The P-CSCF establishes a security association between itself and the IMS terminal. The security association requirements and procedures are provided in 3GPP TS 33.203[82]. The IPSec security associations provide integrity protection. The P-CSCF also authenticates the user and asserts the identity with rest of the nodes in the network, to avoid redundant authentication requirements.

- The P-CSCF forwards the SIP register request from the UE to the appropriate I-CSCF determined from the home domain name provided by the user. This allows for the successful IMS registration of the user/UE.

- The P-CSCF forwards the SIP request or responses to and from the UE to the allotted SIP server, which could be an S-CSCF. The address of the S-CSCF would have been obtained by the P-CSCF as a result of the registration process.

- The P-CSCF performs SIP message compression/decompression for the purpose of reducing the size of the messages and thus reducing the reducing the transmission time and quicker session establishment.

- The P-CSCF may also include a Policy Decision Function (PDF). It performs the task of authorizing the bearer resources and performing QoS management over the media plane. The PDF may of may not be included in the same physical unit. Details about this function of the P-CSCF provided in 3GPP TS 23.207 [83].

- Like the S-CSCF the P-CSCF also generates the CDR and forwards the information to the charging collection node.

An IMS network may have multiple P-CSCF for scalability and redundancy. The P-CSCF may be located in the home network or in a visited network.

### 7.1.1.4 Interrogating-Call Session Control Function (I-CSCF)

The I-CSCF is SIP proxy server; it is placed at the edge of the administrative domain of an IMS network. It is a point of contact for a connection destined to a user who belongs to that network, or a roaming user currently located within the service area of that network operator. The address of the I-CSCF is listed in the DNS (Domain Name System) database and is made available when a SIP server follows the protocol for locating a SIP server for the next hop, the protocol is provided in RFC 3263 [84]. The functions performed by the I-CSCF are mentioned below [75]:

- During the registration process the I-CSCF assigns an S-CSCF for a particular user. The I-CSCF communicates with the HSS and SLF just like the S-CSCF over the Cx and Dx reference points. It uses the DIAMETER protocol. The user information is received by the I-CSCF and depending on the requirements; it assigns an S-CSCF to the user if one is not already allocated.
- The I-CSCF routes a SIP request from another network to the S-CSCF after obtaining the address of the appropriate S-CSCF from the HSS.
- The I-CSCF may encrypt certain parts of the SIP message which may contain sensitive information about the home domain; this functionality is optional and is called THIG (Topology Hiding Inter-network Gateway).
- Like other CSCFs, the I-CSCF also generates CDRs to be transmitted to the charging collection node.

### 7.1.1.5 Application Servers

An application server provides value added services and can be located in the home network or any third party network. It is essentially a SIP sever which hosts and executes various services. There are different modes of operation for an Application Server. It can operate in SIP proxy mode, SIP User Agent mode (terminating or originating), SIP Back-to-Back User Agent (B2BUA) mode etc.

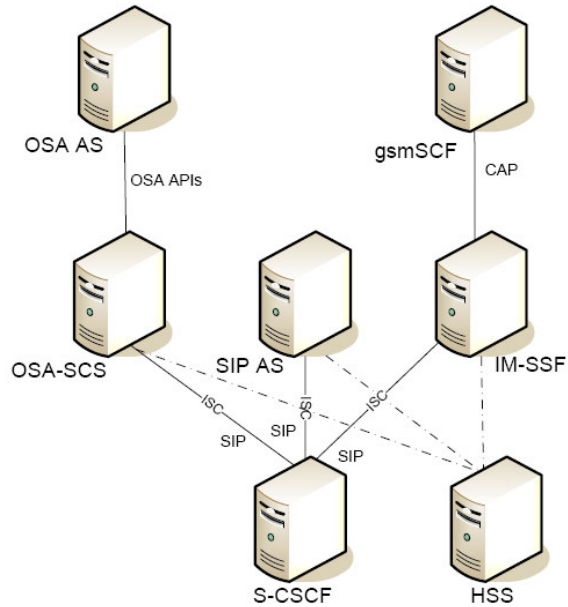

**Figure 5. Types of application servers in IMS**

The S-CSCF interfaces with the AS through the ISC (IP Multimedia Service Control) interface. The ISC interface is based on SIP [79]. Figure 5 shows three different types of Application Servers; they are described below.

- **SIP AS:** The SIP Application server is the native AS. It hosts and executes IP multimedia services based on SIP. All the new services that are going to be developed for the IMS architecture would be implemented using the SIP AS.
- **OSA-SCS:** The Open Source Access-Service Capability Server (OSA-SCS) provides an interface to the OSA framework applications. An AS located at a third party location will not be able to securely connect with the IMS network, whereas OSA has the capability to establish a secure connection with the IMS network. The OSA-SCS inherits all the abilities of OSA and is used to provide secure connectivity for a remotely located AS to the IMS network. The OSA-SCS acts as a regular AS and interfaces with S-CSCF via SIP on one end and as an OSA AS using OSA Application Programming Interface (API) on the other end. The OSA API is described in 3GPP TS 29.198 [85].
- **IM-SSF:** The IP Multimedia Service Switching Function is a specialized application server which allows for integration and reuse of the traditional applications developed for the GSM architecture. CAMEL (Customized Applications for Mobile network Enhanced Logic) was the name of the services that were developed to provide multimedia or enhanced services for GSM handsets. The IM-SSF acts as an application server on one side interfacing with the S-CSCF using SIP, and on the other side it acts as a Service Switching Function (SCF) interfacing with the gsmSCF with a protocol based on CAP (CAMEL Application Part). The CAP protocol is defined in 3GPP 29.278 [86].

The three AS mentioned above may perform different tasks but they behave exactly the same towards the IMS network. They all appear as SIP AS behaving in one of the earlier mentioned modes. The Application Servers present in the home network may optionally interface with the HSS. The SIP AS and the OSA-SCS interface with the HSS using DIAMETER protocol [78] and the interface is labeled 'Sh'. The IM-SSF interfaces with the HSS using protocol based on MAP (Mobile Application Part) defined in 3GPP TS 29.002 [87].

### 7.1.1.6 Breakout Gateway Control Function (BGCF)

As shown in Figure 4, the BGCF provides connectivity to the Circuit Switched domain through the MGCF (Media Gateway Control Function), SGW (Signaling Gateway) and the MGW (Media Gateway). These three nodes put together are referred to as the PSTN/CS Gateway. The BGCF is basically a SIP server which has the additional capability of routing and establishing sessions based on telephone numbers as user addresses. The BGCF is used exclusively for sessions initiated by an IMS user who needs to communicate with a user in the PSTN or PLMN (Public Land Mobile Network) domain, both of which are in the Circuit Switched domain. The main functions performed by the BGCF are as follows:

- It receives a request from the S-CSCF to select the appropriate PSTN/CS Domain break out point for a particular session.
- The BGCF selects the network in which the internetworking with the PSTN/CS Domain is to occur. If the interworking with PSTN/CS domain is to occur in the same domain, it selects the appropriate MGCF and forwards the SIP signaling to it.
- If the interworking with the PSTN/CS domain is to be done at a different network, the BGCF forwards the SIP information to the BGCF of that network. If network hiding is required, the BGCF will forward the SIP signaling through the I-CSCF to the other BGCF.
- The BGCF also generates CDRs to forward to the charging collecting node.

### 7.1.1.7 Public Switched Telephone Network/ Circuit Switched (PSTN/CS) Gateway

The PSTN/CS gateway comprises of three components as mentioned earlier. The functions performed by each of them are given below.

**MGCF (Media Gateway Control Function)** interfaces with the BGCF and receives the SIP signaling. Its function is to convert the SIP signaling to either ISUP (Signaling System 7) defined in ITU-T Recommendation Q.761 [88] over IP or BICC (Bearer Independent Call Control) defined in ITU-T Recommendation Q.1901 [89] over IP. The converted signaling is forwarded to the Signaling Gateway (SGW). The MGCF also controls the resources in the Media Gateway (MGW). The MGCF and the MGW communicate with the help of the H.248 [90] protocol, specified in the ITU-T Recommendation H.248.

**SGW (Signaling Gateway)** provides the signaling interface with the circuit switched domain. Its main function is to perform lower level protocol conversion. It converts MTP (Message Transfer Part) defined in ITU-T Recommendation Q.701 [91] into SCTP (Stream Control Transmission Protocol) defined in RFC 2960 [92] over IP. So the signaling format ISUP or BICC over MTP is transformed into ISUP or BICC over SCTP/IP.

**MGW (Media Gateway)** connects the media plane of the PSTN or any other CS environment with the media plane of IMS. The MGW transcodes the IMS data transported over RTP (Real Time Protocol) defined in RFC 3550 [93] into PCM (Pulse Code Modulation) used in the PSTN environment. Also the MGW performs transcoding in situations where the IMS terminal does not support the codec being used by the CS side.

### 7.1.1.8 Media Resource Functions (MRFs)

The Media Resource Function (MRF) handles all the media transportation and processing requirements. It is divided into two functional components as shown in Figure 4, the Media Resource Function Controller (MRFC) and the Media Resource Function Processor (MRFC). The MRFC interfaces with the S-CSCF over the Mr interface and uses SIP [57] for signaling purposes. The tasks performed by the MRFC are as below [80]:

- It controls the media stream resources in the MRFP.

- The MRFC interprets the information forwarded by the S-CSCF and the Application Servers and modifies the operation of the MRFP according to the directions.
- The MRFC generates CDRs like the other nodes in IMS to be forwarded to the charging collecting node.

The MRFP is controlled by the MRFC though the Mp interface, also called a reference point. The Mp reference point does not have a specific protocol specified for it yet and has an open architecture to allow extension work to be carried out. It completely supports the H.248 Standard [90]. The tasks performed by the MRFP are given below [61]:

- The MRFP controls the bearer plane on the Mb reference point.
- It provides the functionality of mixing various incoming media streams in case of a conference call.
- It acts as a source of media streams or plays streams as for multimedia announcements.
- The MRFC performs all other media processing functions such as transcoding, media analysis etc.
- It also provides floor control or manages access rights in a conference environment.

## 7.2 IMS Protocols

The protocols used in the IMS environment are derived from the internet and wireless domain (GSM/GPRS). 3GPP decided to use the protocols being developed by the IETF and ITU-T for the IMS and thus was able to capitalize on their expertise in designing robust protocols. A brief description of the various protocols used in IMS is given below.

### 7.2.1 Session Control in IMS

The protocol used for session initiation and control in IMS over IP networks is the Session Initiation Protocol (SIP) specified by the IETF. SIP is a text based protocol unlike other session protocols such as BICC and H.323. This makes it easier to debug, extend, and build services on it. One of big reasons for choosing SIP was the fact that it is based on many familiar and successful protocols such as SMTP (Simple Mail Transfer Protocol) and HTTP (Hypertext Transfer Protocol). Also SIP follows the familiar client-server model. SIP makes is very easy to develop new applications, which is one of the requirements of IMS.

### 7.2.2 Authentication, Authorization and Accounting (AAA) in IMS

Authentication Authorization and Accounting (AAA) operations play a very important role in any network, especially in the IMS environment. It is of great importance to have an efficient and highly reliable mechanism to perform the tasks of authenticating a user's identity, authorizing the user to access the appropriate resources and making sure the resources and services consumed are logged accurately and billed correctly. IMS uses the DIAMETER protocol to perform the AAA operations. It allows different nodes to access retrieve or modify user information from HSS or SLF. The DIAMETER protocol is an improvement over the older RADIUS protocol [94]. The DIAMETER protocol is defined in RFC 3588 [60]. The DIAMETER protocol is used over different interfaces such as Cx, Dx and Sh. It consists of a base protocol and is used to develop various DIAMETER applications. These applications are extended and customized for a particular purpose or an environment. The different interfaces may use different DIAMETER applications to perform the various AAA procedures.

### 7.2.3 Quality of Service in IMS

Generally there are two models to provide QoS on the packet switched IP domain specifically the Internet. They are the Integrated Service model and the Differentiated Service (DiffServ) model.

Integrated Service model is defined in RFC 1633 [83], it provides end to end QoS. The endpoints request a certain QoS and the network grants it. The protocol used by the Integrated Services architecture is RSVP (Resource reSerVation Protocol), it is specified in RFC 2205 [95] and has been updated by RFC 2750 [96] and RFC 3936 [97]. Integrated Service works well in small networks does not scale well as the routers have to store state information about every flow and perform lookup before routing any packet. The second model of QoS solves some of the problems faced while using the Integrated Service model. The DiffServ architecture is specified in RFC 2475 [98] and RFC 3260 [99]. The DiffServ servers need to maintain minimum state information about the flows and enables a quicker treatment for the packets flowing through them. In this architecture the router is aware of the treatment that needs to be given to each packet; the treatment is referred to as the Per Hop Behavior (PHB). Each PHB is identified by 8-bit codes called Differentiated Service Code Points (DSCP). The DSCP information is carried by the packets in their IP headers. In IPv4 it is placed in the 'Type of Service' field and in IPv6 it is placed in the 'Traffic Class' field.

IMS allows many different end-to-end QoS models. All the models are described in 3GPP TS 23.207 [83]. The terminals may use link layer resource reservations methods such as PDP context reservation, or directly use protocols such as DiffServ or RSVP. The IMS networks use DiffServ and may use RSVP.

### 7.2.4 Security in IMS

Security generally deals with integrity, confidentiality, and availability. There are various means to achieve the security requirements in the SIP environment. In IMS security can be divided in two different areas, Access security and Network security.

Assess security deals with authentication and authorization processes and establishment of the IPsec security authorization (architecture defined in RFC 2401 [100]); these procedures are performed during the REGISTER transaction. All the procedures for security access are provided in 3GPP TS 33.203 [101]. In the 3GPP networks the user identity is stored on a smart card inserted in the IMS terminal; this card is usually known as UICC (Universal Integrated Circuit Card).

Network security deals with protecting the traffic between two nodes. The nodes may or may not belong to the network. There may be different levels of requirements from the network security mechanisms in place. If two different security domains are involved, the traffic travels through two Security Gateways (SEG). In this case the traffic is protected using IPsec ESP (Encapsulated Security Payload), specified in RFC 2406 [102] and runs in tunnel mode. The security associations are established and maintained using IKE (Internet Key Exchange), specified in RFC 2409 [103]. All the network security requirements are mentioned in 3GPP TS 33.210 [104].

### 7.2.5 Policy Control in IMS

Policy control deals with the media-level access control; the decisions made by the policy control mechanism authorize a user to use the media plane and assigns the QoS to be provided for that user session. The media-level policy is enforced by the routers present in the network, but these routers do not have the ability to make decisions about users as they do not have access to the user information stored in the HSS. The task of obtaining the user information and making these decisions is performed by a SIP server in this case. The SIP server informs the routers to allow of deny a certain user with the requested media resources.

The node which makes the decision, in this case the SIP Server is called the Policy Decision Point (PDP) and the router is called the Policy Enforcement Point (PEP). The protocol used between the PDP and PEP is called Common Open Policy Service (COPS) protocol, it is defined in RFC 2748 [105] and has been updated by RFC 4261[106], which provides a higher level of security at the transport level. COPS generally supports two models for policy control, the outsourcing model and the configuration model (also called provisioning).

In the outsourcing model the PEP contacts the PDP for every decision, whereas in the configuration model the PEP stores the policy from the PDP locally and uses it to make decisions. IMS uses a combination of the two models; it's called COPS-PR. It's a mixture of the two models as it uses the same message format and the Policy Information Bases (PIB) as used by the provisioning model and the policy decision are transferred in real time like in the outsourcing model.

In IMS there are two types of limitations on the session that can be established. They are user-specific limitations and general network related policies. The user-specific limitations include restrictions on a particular user, in terms of resources that are allowed. An example would be an audio only subscription, so the user will not be allowed to establish video sessions. The general network policies would apply to all the users of that network. This might include restrictions on the codecs that can be used. The P-CSCF deals only with the enforcement of the general network policies, whereas the S-CSCF handles both user-specific policies and the network policies. Both these PDPs use the same mechanism to monitor the sessions. They access the SDP (Session Description Protocol) body to identify the type of session and media requested during SIP procedures.

# 8   Convergence using IMS - A Case Study

In this section, we present an illustrative case study on convergence using IMS in a residential environment. First we describe the requirement of a converged residential environment.

Bundling up of communication services has been very successful of late. Companies have been providing Internet, voice telephony, and digital entertainment services to residential subscribers with the convenience of a single bill. These services include cellular telephony also. What lacks in this environment is interworking among various services.

A converged environment would allow for rich multimedia applications to be accessed by the users on any device and retain the user profile and other settings. The idea is to be able to communicate, establish multimedia sessions, and perform control and configuration operations on all the networked elements in the residential environment. The following are the typical requirements of a converged residential network:

- The system should allow multiple user profiles with different levels of control over services and the systems.
- It should also allow integration of wireline and wireless (cellular) voice communication, or should allow the users to make and receive phone calls on wireline-based digital/analog phone and the wireless cellular phone interchangeably as per personal preference.
- The system should allow for multimedia sessions to be established between the various terminals (audio/video database server, TiVo etc) in the network. The users should be able to access the content within the residential environment on any terminal or from outside the residential environment (restricted only by the capabilities of the device and the available bandwidth).
- The nodes in the residential environment may use either wired or wireless connectivity. The local area network should support secure wired and wireless access (WiFi or WiMAX).
- The network must support application or service hosting and the application server should be remotely manageable. These applications may include but are not limited to residential security system (monitoring, configuring and authorization), web hosting services, residential power/gas monitoring and control etc.
- The residential environment should be secure, easy to manage, and efficient, and should provide effective means of managing and configuring devices in the network.

Figure 6 shows such a converged architecture which shows a few of the many possible elements in the network. All these elements should be accessible to the user with sufficient rights, both locally and remotely. The users in the residential network subscribe to a number of services which might include cable television (digital or analog), Internet access, telephone connection, and one or more cellular phone subscriptions. As shown in the Figure 6, we will assume these services are being provided by the Multimedia/Communication service providers. All the elements in the residential environment are networked as in a LAN environment. All there services are controlled by a so called residential server which performs the control operation and a wireless router which provides the connectivity. There could be other systems operational in the residence, such as a security system, an electronic power supply and control mechanism, which controls the lights, air conditioning and other energy related functions, etc. All these are linked and controlled by the residential server as well.

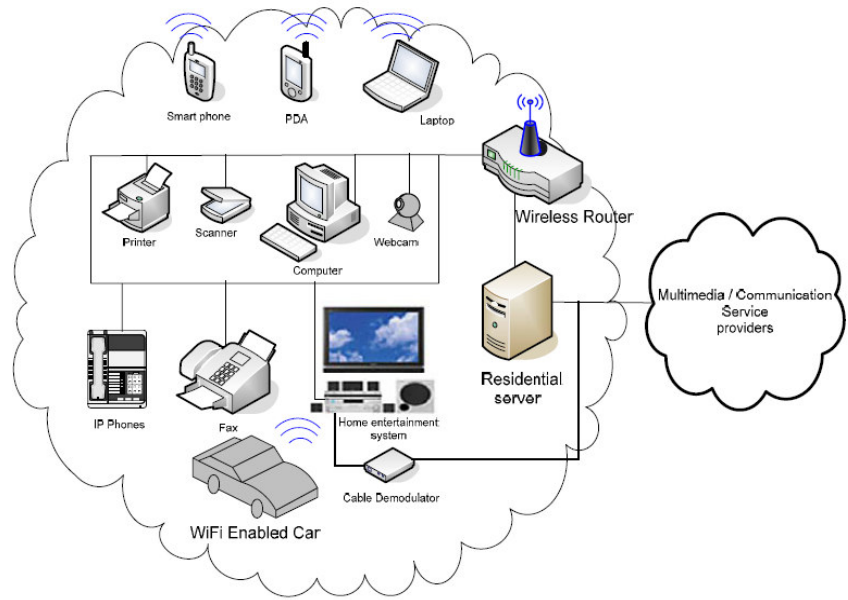

**Figure 6. Converged home network environment**

## 8.1 Scenarios in converged residential network

We shall consider a family comprising four people, two parents and two children. Each of these four users might need different services and will have different levels of authority over the system. Let's consider a problem or a convergence requirement of this residential network and work on providing a solution for it. We assume that say the father (John) has a cell phone (GSM/UMTS). He also subscribes to digital television services from the same company which provides him with a digital telephone and Internet assess. This is a typical scenario at present for a customer of triple play services.

What John desires from this converged system is to have the following features:

- When at home, he should be able to make and receive calls (voice and maybe video) using his home digital phone or cell phone interchangeably. This means that once he gets home his cell phone calls should be automatically routed to his home phone to save minutes.

- The home phone is connected to a high bandwidth connection; it's an IP phone and should be allowed to be implemented over the WiFi network. The cell phone used by John may be a smart phone which has WiFi connectivity, so he can continue to use his handset to receive calls intended for both his home phone number and cell phone number but with enhanced bandwidth, improved display and reduced cost of access.

-  Other members of the house may or may not have a cell phone account; if they do same should apply for them. Also the system has to be smart enough to identify the called party and not alert John on his phone if the call is for his children. This needs to work both ways as he does not need to his calls to be forwarded to other members of his family. The system should be configurable to manage this.

- He should be able to continue to access the internet from any other device, including a desktop, a laptop, a PDA/Smart phone etc. Also he should be able to receive communication (voice call, email, voice mail, instant message, video call, etc.) addressed to his various accounts (email, home phone number, cell phone number, etc.) on any of the devices listed above. They have different abilities in terms of processing power, screen size, etc.

- The entertainment services being subscribed by the family may include digital TV, access to other online multimedia services, music, videos etc. The family must have access to the entertainment services from any network node capable of playing audio and video. This includes watching a particular TV channel on a PC, a laptop or a PDA.

- The family might like to share some data among themselves and their friends; this might be photos, video, audio or other data. They should be able to do that securely, from within the home and even outside home. John might use a database server which would be accessible to all authorized people.

- John would like to be able to add services or applications being run in the home network without changing the system much and should be able to remotely configure and control those services. These might include a security system, a web server, etc.

The requirements mentioned above are quite advanced. However, the NGN system needs to be designed that should be able to handle all these requirements and more. In the following section, we propose a mechanism of establishing such a system based on enhanced IMS architecture.

## 8.2 Achieving converged residential network using IMS

There are certain pre-requisites for establishing a session in the IMS environment in the residential network described in the previous section. We first present these re-requisites:

- **Establishing an IMS service contract:** This includes establishing a subscription with the IMS service provider. During this process, the service provider will provide the customer with the appropriate identities and the service profiles will be created depending on the user's requirements in terms of services, bandwidth for those services, and access to various other applications being provided by the service provider. After the service contract is established, the user profile will be stored in the HSS and will be used during various operations in IMS, including Authentication, Authorization and Accounting purposes etc.

- **Obtaining and IP address:** Every IS terminal needs to get connected to the IMS core network. The connectivity is provided by the IP-CAN (IP-Connectivity Access Network). This could be any IP-based transport network such as GPRS (as in GSM/UMTS network), xDSL, Wireless LAN through Wi-Fi (IEEE 802.11) or WiMAX (IEEE 802.16) networks etc. IMS uses only IPv6 address.

- **Discovery of P-CSCF:** After the IMS terminal obtains IPv6 address, the next step is to locate a P-CSCF. This procedure includes the discovery of the IP address of the P-CSCF, which acts as an inbound/outbound SIP proxy server and will interact with the IMS core network.

- **IMS registration:** The registration process in IMS is based on the SIP registration process, where a public user identity is bound to a SIP URI (Uniform Resource Identifier). This SIP URI contains the IPv6 address or the host name of the terminal where the user is reachable. This is done using the SIP REGISTER request defined in RFC 3261 [79]. The registration process in IMS needs to accomplish the following tasks:

- o Bind a Public User Identity to a contact address
- o Authentication of the user by the home network
- o Authentication of the network by the user
- o The SIP registration is authorized by the home network and allows for the usage of the IMS resources subscribed by the user
- o Verification of a roaming agreement between the home network and the visited network is the P-CSCF is located outside the home network, thus authorizing the usage of resources
- o The home network informing the user about the various other identities that have been allocated to the user (implicitly registered user identities)
- o Negotiation of the security mechanisms between the IMS terminal and the P-CSCF for subsequent signalling and other such security association to protect the integrity of the messages
- o Uploading of the compression algorithms between the IMS terminal and the P-CSCF.

**Assumptions:**

- The users in the residential network are IMS subscribers and are assigned Public and Private User Identities by the IMS provider. The IMS provider provides the users with a UICC (Universal Integrated Circuit Card), which may contain an ISIM (IMS SIM) application, or an UMTS (UMTS SIM) application or both. The information stored on this card includes among other things are the Private User Identity, and one or more Public User Identities.
- The IP Connectivity Access Network (IP-CAN) used by the residential network is operated by the IMS provider.
- The residential network is an IP-based Ethernet environment, and uses IPv6 protocol.
- All the networked devices have IPv6 interfaces (NIC) and can obtain a globally unique IPv6 address.
- No traditional PSTN terminals are being used and all terminals are IP-based. However, the IMS terminals and users are fully capable of calling PSTN or other CS customers and vice versa.
- The entertainment services are digital and use configurable IP-based set-top devices.

**IMS Implementation of the residential network**

Figure 7 depicts the proposed architecture for the residential network. Since the IMS service provider also provides the IP-CAN, all the nodes in the residential environment can be considered to be a part of the Home (IMS provider's) network. The location of the P-CSCF and the SIP Application Server (SIP AS) in unorthodox in the network shown in Figure 7. However, they are technically still in the home network of the IMS provider. The architecture allows for a high level of flexibility in terms of the IMS service provider. The service provider can be changed without any major changes in the network, except for some reconfiguration of the nodes.

The IMS agreement is established to allow the user (in this example John) to register any of the public user identities at any of the available IMS terminals. The abilities of the terminal and the subscription of John regulate the type of sessions that can be established. The IMS terminals obtain an IPv6 address from the IP-CAN provider using the required DHCPv6 and DNS procedures.

The P-CSCF allotment is also regulated by the IMS service provider. In this case, we assume that the service provider has located an outbound SIP proxy (P-CSCF) at the customer's premises and this P-CSCF is allocated to any sessions established by the user from the residential network. This may be done by configuring the DHCPv6 for a certain set of addresses or terminal names.

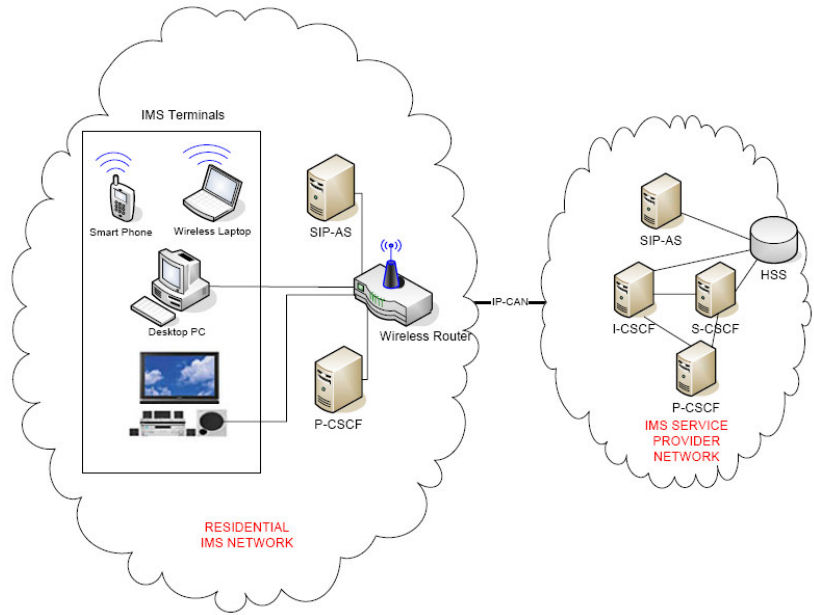

**Figure 7. The proposed IMS implementation in the residential network**

In the registration process, the user information stored in the USIM or ISIM (present in the UICC-Universal Integrated Circuit Card) would need to be accessed and transmitted in the right format from the IMS terminal. To accomplish this task, the IMS terminal being used would need to be connected in some way with the smart card allocated to the user. This smart card would probably be present in the wireless handheld device (probably a smart phone) being used by the user, and the user might perform this task by docking the smart phone using a cable or running a secure wireless application which transmits the relevant information from the smart card to the IMS terminal used. Once the authentication and authorization is complete and the S-CSCF is allocated to the user, it goes through the process of evaluating the initial filter criterion, which may result in invocation of one or more Application Servers.

Suppose, we need to provide convergence between the wireless cell phone used by John and the residential digital phone present in the residential network. We assume that there are two service profiles being used by John. Our task at hand when John is registered under the *home service profile* is to handle all incoming calls (voice, video or multimedia) to John's *general service profile* as per John's directive. He may wish for calls from certain people to be forwarded to his Home contact address (his residential number) and the rest of them might be forwarded to some sort of automated response and messaging system (answering machine); there are various other possibilities or ways in which John wishes to be reached while he is registered using his home profile.

This is achieved using the *initial filter criterion* in present in John's user profile. Let us assume that John maintains a list of people (with known contacts) from whom he wishes to establish incoming communication no matter where he is registered (home or outside). This list is stored in one of the application servers, say *AS1*. Another Application Server, say *AS2*, receives input from AS1 regarding the session and acts as a SIP proxy and directs the session to the appropriate location. Both these Application Servers can be configured by the user over the 'Ut' interface--the interface is between the User Equipment and the Application Server and is used exclusively for the purpose of configuring access related information and not for live traffic. The security functions for this interface are defined in 3GPP TS 33.222 [107].

When John registers at home, the S-CSCF invokes the *AS1* which has been configured by John as per his wishes and regulates his presence at home. This information is used as an input to *AS2* which receives the contact address of John and *AS1*. *AS2* will also have access to the list of people who would be permitted to contact John at home.

# 9 Convergence Standardization Organizations

The next generation networks (NGN) or advanced networks and systems are being developed in different parts of the world and by various agencies and organizations. Most of them are working in tandem to develop a uniform standardized system. The purpose of this section is to clearly identify the reasons for these efforts and their objectives. We will examine the requirements of the NGN systems as defined by ITU and will look at the various organizations working together to obtain a standardized solution for the same. As mentioned earlier, there are a number of agencies working towards the goal of advanced future networks. A major contributor in this direction is 3GPP and 3GPP2, which have introduced IMS. In the USA, ATIS (Alliance for Telecommunication Industry Solutions) is leading the way in developing NGN systems based on extended IMS. We will look into their operations in some depth.

The objective of ATIS, and more specifically ATIS Next Generation Network- Focus Group (NGN-FG) is to design NGN which has the following fundamental aspects:

- Packet-based transfer
- Separation of control functions among bearer capabilities, call/session, and application/service.
- Decoupling of service provision from network, and provision of open interfaces
- Support for a wide range of services, applications and mechanisms based on service building blocks (including real time/streaming/non-real time services and multimedia)
- Broadband capabilities with end-to-end QoS and transparency
- Interworking with legacy networks via open interfaces
- Generalized mobility
- Unrestricted access by users to different service providers
- A variety of identification schemes which can be resolved to IP addresses for the purposes of routing in IP networks
- Unified service characteristics for the same service as perceived by the user
- Converged services between fixed/mobile
- Independence of service-related functions from underlying transport technologies
- Compliant with all regulatory requirements, for example, concerning emergency communications and security/privacy, etc.

ATIS NGN-FG is driven by the business needs of the North American market; the aim is to produce as much as possible international NGN standards. For this reason, it works in collaboration with a number of global standardization agencies.

Figure 8 shows the interaction of ATIS with other organizations and agencies involved in the standardization process of next generation converged networks.

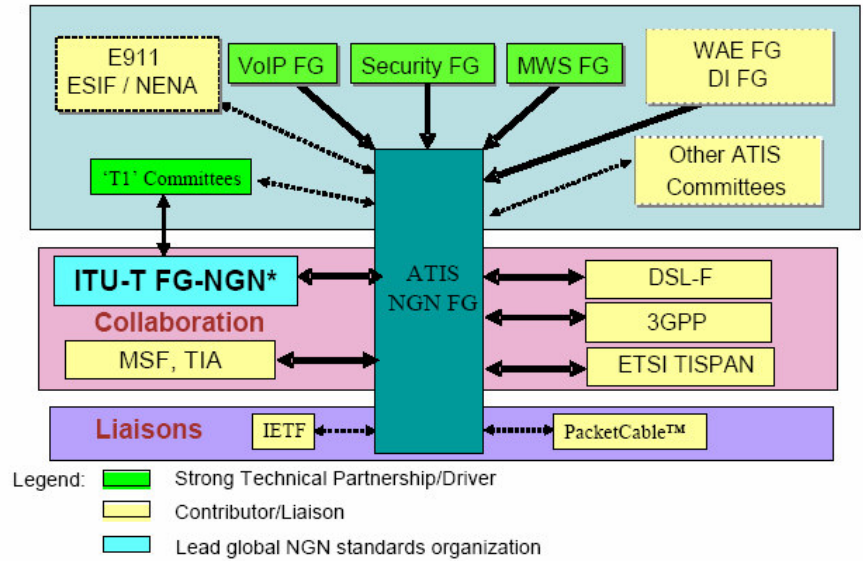

**Figure 8. ATIS standards collaborations**

Source: ATIS Next Generation Network (NGN) Framework, Part I: NGN Definitions, Requirements, and Architecture, Issue 1.0, November 2004, pg. 11

The list of partners in Figure 8 is given below:

- ITU-T: SG13, including the Focus Group on NGN (FGNGN)
- 3$^{rd}$ Generation Partnership Projects (3GPP)
- European Telecommunication Standards Institute (ETSI) TISPAN
- Multiservice Switching Forum (MSF)
- DSL Forum (DSL-F)
- CableLabs
- Institute for Electrical and Electronics Engineer (IEEE)
- ATIS Technical Committees (e.g., PTSC, TMOC, NIPP)
- 3$^{rd}$ Generation Partnership Project #2 (3GPP2)
- Telecommunications Industry Association (TIA)
- Internet Engineering Task Force
- National Emergency Numbering Association (NENA)
- Emergency Services Interconnection Forum (ESIF)
- TTY Forum
- Industry Numbering Committee (INC)
- FCC Network Reliability & Interoperability Council (NRIC)
- Open Mobile Alliance (OMA)
- Metro Ethernet Forum (MEF)
- MPLS and Frame Relay Alliance

ITU-T study groups including SG-13 and its Focus Group on NGN (FGNGN) are responsible for the global standards for NGN telecommunication systems. The idea is to expedite the process of standards development by having regional Standard Development Organization (SDO) submit contributions to the IT-T. Such a workflow provides and establishes a global scope to the efforts of ITU-T and allows for a more harmonized approach.

TISPAN is the ETSI group responsible for all aspects of standardization for present and future converged networks, including the NGN and including service aspects, architectural aspects, protocol aspects, QoS studies, security related studies, and mobility aspects within fixed networks, using existing and emerging technologies. TISPAN has developed an architecture for NGN based on 3GPP IMS subsystem. This architecture is named extended IMS and is considered as the base for ATIS NGN work.

There are many other organizations and agencies which ATIS partners with to achieve its goals and objectives. Different agencies define standards for different mediums of communications, so it is essential for these organizations to interact and develop interoperable standards. For example, DSL Forum defines standards for data communication over the telephone lines, CableLabs develops standards for communication over the cable infrastructure and IEEE has been developing and defining standards for the data communication for the LAN/MAN, in both wired (IEEE 802.3) and wireless medium (IEEE 802.11-WiFi and 802.16-WiMAX).

## 9.1 Alliance for Telecommunications Industry Solutions (ATIS)

The ATIS NGN Framework documents Parts I and II provide a detailed set of requirements for the NGN systems. These documents examine every aspect of telecommunications and clearly define expectations from the NGN systems being developed. The requirements have been divided into six major groups:

- General requirements
- US regulatory requirements
- End user applications
- Network service enablers
- Underlying network/support capabilities
- Business model-driven requirements

The general requirements section [108] of the document contains guidelines about the basic functionality that would be required to exist for the NGN to fulfill the ITU-T requirements. It includes issues such as NGN network interconnection requirements, different types of interfaces between the *application service providers* (ASP) and the *next generation service providers* (NGSP), mechanisms to measure and predict QoS, guidelines for incremental replacement of legacy services, PSTN simulation, PSTN emulation, mobile network evolution, transparent end-to-end communication, synchronization and timing issues, etc.

The US regulatory requirements section of the document lays down all the possible regulatory requirements that might be enforced on the NGN systems. The regulatory requirements are a moving target and may evolve with the regulatory and legislative actions of the various government agencies at different points of time. These requirements include information and guidelines regarding various service-specific or general regulations such as *lawfully authorized electronic surveillance* (LAES), number portability regulations, number pooling, E 9-1-1 directives, *emergency telecoms service* (ETS), FCC rules and regulations, accounting etc.

The end user applications section of the document provides guidelines regarding the user applications and about the way they would need to be approached in the new NGN network architecture. The NGN would require application to be supported on a common, converged architecture, so there would be need for changes or variations in the services and their inherent capabilities. This section lays down requirements for applications such as interactive voice, content and capabilities of video services, multimedia conferencing, content sharing, basic and advanced interactive gaming, sensor and control networking, mobility management etc.

Some of the key *network service enablers* are defined in the next section of the document. Even though the NGN is not vertically integrated, there are still some network-based communication services that bring value to customer applications. These include QoS requirements, presence services, policy definitions and enforcement, media resource and media gateway functions, personal profiles unified interface, and service ubiquity. There are other important aspects which have been defined and described in this document such as roaming, location services, personal information management and access, digital rights management, and session management.

The underlying network/support capabilities are those that are not directly accessible by the applications. These are of monumental importance and need to be introduced in the core network and are very well-defined in this section of the document. These capabilities include *operations administration maintenance and provisioning* (OAMP), security provision, integrity requirements, confidentiality and privacy requirements, attack mitigation and prevention policies, accounting (ordering and billing), trust policies are requirements etc.

The last segment of the requirements identifies and defines various business model-driven requirements. This section covers topics such as the operational expense (OPEX), implications for service providers, third part access implications, service delivery environment, and consolidated operations requirements.

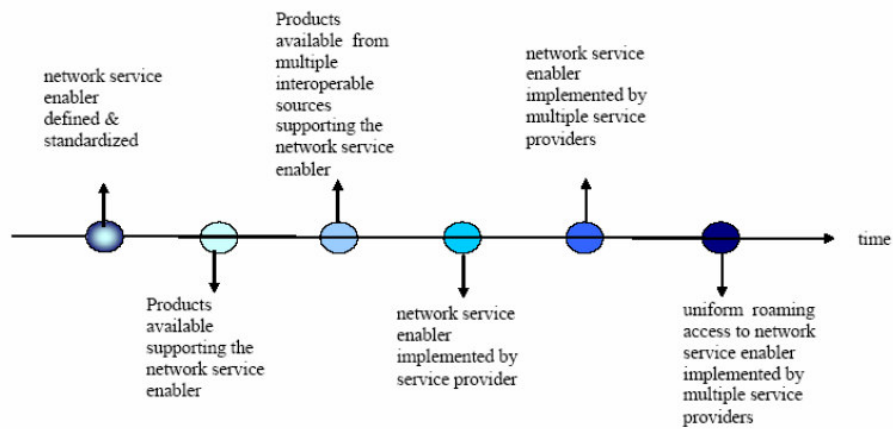

**Figure 9. NGN capability deployment roadmap**

The approach adopted by ATIS to develop NGN systems includes identifying a roadmap. The primary purpose of identifying the NGN roadmap is the creation of an infrastructure that enables flexible and efficient creation of NGN services. The focus of the roadmap is to identify the underlying network service enablers that will allow potential new services to be introduced. The roadmap is developed keeping in mind the heterogeneity in the industry in terms of technological (IP network capabilities) starting positions in moving towards NGN. The network service enablers required for developing new services suggested as the starting point of the NGN include the following:

- Unified user profile

- Security (various aspects as defined in NGN framework)

- Decoupling of services from access technologies

- Integrated management of all services, users and networks

- Presence
- Scalable management and operations
- Quality of Service (QoS)
- Settlement (accounting)
- Digital Rights Management (DRM)
- Media Resource Functions (MRF) etc.

All these network service enablers have been assigned different priorities and have other functions dependent on them [109]. Figure 9 illustrates the flow of new network service enablers.

## 9.2 Internet Engineering Task Force (IETF)

As IETF is an important standardization body for Internet protocols and since it is playing an important role in evolution of IMS standards, we provide a very brief description about its activities. IETF is a loosely organized, self governed organization consisting of a wide variety of people with different technical backgrounds including network designers, vendors, etc. Their efforts are directed towards the development of the architecture, protocols, and the operations of the public Internet. Their mission statement is documented in RFC 3935 [110].

The IETF is organized into a number of working groups and the actual technical developmental work is done under one of the working groups. These working groups are organized into Area Directorates, and all these are managed by the Internet Engineering Steering Group (IESG). The IESG is responsible for the technical management of the IETF and decides the area the IETF should work on. The members of the IESG also review all the specifications that are produced.

The technical documents used within the group are called Internet Drafts. There can be two types of drafts: individual submission and working group items. The individual submissions are reviewed by the members and become group item if they are found worthy of investigation by the rest of the group. Work is done on the group item and is eventually submitted to the IESG, when the group is confident that it's ready for publication. The IESG provides feedback and approves the publication of a new RFC (Request for Comment). The internet drafts can be considered to be stable specifications only after they are RFCs.

## 9.3 Third Generation Partnership Project (3GPP)

3GPP and 3GPP2 are standardization bodies responsible for the development of the 3G cellular protocols and standards. Currently 3GPP is involved developing the long term evolution (LTE) standards for next-generation cellular communication networks. 3GPP comprises a number of international *standard development organizations* (SDO) such as ARIB, ETSI etc. 3GPP is organized into *technical specification groups* (TSG) and they are managed by a supervising organization called the *project co-ordination group* (PCG). The TSGs do not produce standards but they deliver *technical specifications* (TS) and *technical reports* (TR). Once these are approved by the TSGs, they are submitted to the organizational partners for the documents to go through their individual standardization processes.

3GPP2 has a very similar structure and operates in pretty much the same way. 3GPP2 version of IMS is called Multimedia Domain (MMD). 3GPP2 is organized into 4 TSGs. TSG-A deals with *access network interface*. TSG-C focuses on CDMA technology, TSG-S on service and systems aspects, and TSG-X on intersystem operations.3GPP2 also delivers technical specification (TS) and technical reports (TR), but follows a different numbering pattern.

Almost all the protocols selected for IMS were originally from the Internet domain. However, there was an obvious need to modify the existing protocols to meet the requirements of IMS. For this purpose, IETF started jointly developing the protocols with 3GPP/3GPP2. A collaboration was established between these organizations and documented in RFC 3313 [111] (IETF-3GPP) and RFC 3131 [112] (IETF-3GPP2). The focus of collaboration in the Internet area was in the fields of IPv6 and DNS. One of the outcomes of these efforts was the development of RFC 3316 [55] by the IPv6 working group. This RFC provides guidelines for the implementation of IPv6 on cellular hosts. It allows the terminal to recognize a GPRS network and use IPv6.

In the operations and management area the collaboration between 3GPP and the IETF was in the development and modification of the COPS and DIAMETER protocols. A modified version of COPS was chosen to be used in the IMS called COPS-PR. The DIAMETER protocol is used as the base protocol and applications are developed on it. These applications and command codes are provided by the IETF in RFC 3589.

In the transport area most of the work done is for the development of the *session initiation protocol* (SIP). RFC 3261 was an outcome of this effort. However, this was not sufficient as the requirements of 3GPP were not being met. To accomplish this, IETF created a new working group called SIPPING, which collected and prioritized the SIP requirements and forwarded them to the SIP working group, where the actual work was done. This process of collaborative effort to develop SIP is documented in RFC 3427.

# 10 Indian Telecommunication Industry

In this section, a brief description is given of the current Indian telecommunication industry. The emerging trends of the industry are analyzed and the future trend is also presented. Some of the potential challenges that may act as obstacles to growth of this industry and the next generation network deployment in India are also discussed.

India is one of the fastest growing wireless markets in the world. It is now the second-largest telecom market, just after China. Despite the global economic slowdown, the Indian telecom industry continues to grow substantially, delivering strong returns on investment, fuelled by the growth in the wireless industry. The wireless subscriber base grew at a CAGR of 61% over FY04- FY09, while fixed-line subscribers dropped to 38 million in FY09 from 40.9 million in FY04. The subscriber base reached 392 million as of March 2009 with more than 10 million subscribers being added every month. However, India accounts for 7% of the total subscribers in the Asia Pacific. Thus, low mobile penetration provides huge growth potential [114].

Moreover, the teledensity levels between the urban and rural areas vary widely, which suggests untapped potential in the rural segment. It appears quite clearly that the rural areas would be the next growth driver for the Indian telecom industry. However, operators would face certain challenges such as high operating and maintenance costs and low ARPUs and MOU.

A continued decline in tariffs and disruptive entry pricing by new players pose risks to the margins of most of the companies. The wireless Average Revenue Per User (ARPU) is expected to fall further due to increased competition. Also, incremental subscribers are now mainly from the rural areas with lower usage, which should reduce the elasticity of MOU with the decline in tariffs. Increased revenue is expected from non-voice services to support revenue growth.

## 10.1 Current Trends

**Subscriber base:** The Indian telecom market is characterized by low penetration and huge growth potential. The wireless subscriber base grew at a 61% CAGR over FY04-FY09 while fixed-line subscribers dropped to 38 million in FY09 from 40.9 million in FY04. Since wireless penetration is ~30% at present, there is a huge growth potential.

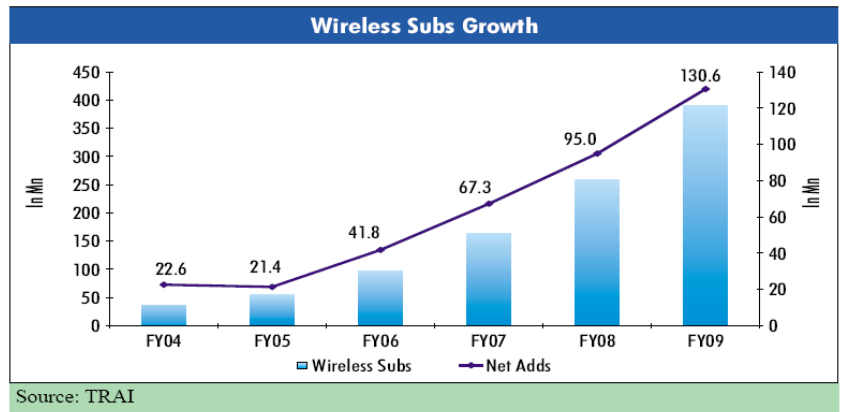

**Figure 10. Wireless subscriber growth and net additions**

**MOU growth not in line with fall in tariff:** The Average Revenue Per User (ARPU) for GSM operators dropped to INR220/month in December 2008 from INR316/month in December 2006, an average drop of 4.4%. Despite this, GSM operators' revenue grew an average 6.2% during the same period. This indicates that subscriber growth and higher Minutes of Use (MOU) per subscriber have more than offset the decline in ARPU. ARPU has been declining sharply as a result of intense competition among operators, fall in tariffs, regulatory policies such as phasing out of Access Deficit Charge (ADC), and reduction in termination charges. On the other hand, MOU has not increased in line with the fall in tariffs, mainly due to low usage of incremental subscribers.

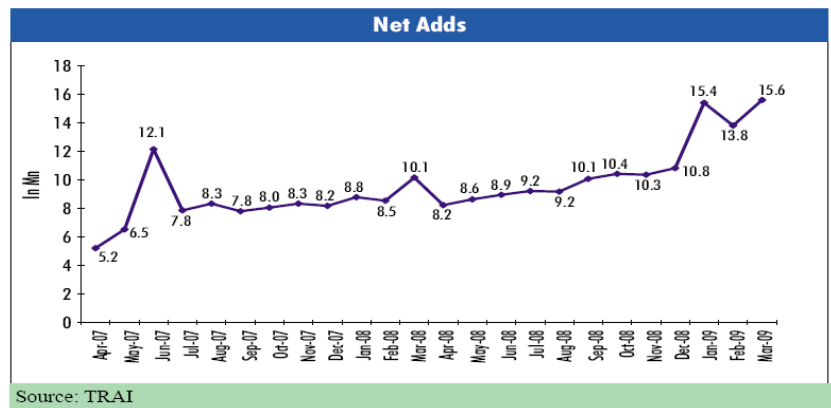

**Figure 11. Monthly net additions of wireless subscribers**

**Rapid growth in investments:** The boom in the domestic telecom market has been attracting huge investments, which will likely accelerate with the launch of new services such as 3G and the entry of new players. The telecom industry saw 145% growth in FDI to INR108 billion over April-January 2009, compared with INR43.96 billion in the year-ago period (Source: DIPP). The pace of investments is expected to accelerate further with the entry of new license holders as well as the auction of 3G spectrum. Furthermore, buoyed by the rapid surge in the subscriber base, the domestic players too are planning huge investments for FY10.

**Industry moving towards consolidation:** The Indian telecom industry saw numerous mergers and acquisitions in the past. Many foreign players have already entered the industry by acquiring stakes in Indian players, who have acquired licenses recently. The 3G auctions would also lead to many foreign players entering the sector. However, as DOT has stipulated that for bidding for 3G spectrum an entity must hold a Universal Access Service (UAS) license, many foreign players might tie up with existing UAS licensees, which in turn could result in mergers.

## 10.2 Future Trends

**Subscriber growth:** The rural subscriber base has been growing substantially due to greater focus of the telecom service providers on rural operations. The rural subscribers accounted for almost 30% of the total subscriber base as of December 2008. However, there is an untapped potential in rural areas (where almost 70% of the population resides), as rural teledensity is quite low at 12.6% vs. urban teledensity of 81.3%. Quarterly rural subscriber additions have begun to exceed urban subscriber additions. Also, growth in rural subscribers has exceeded that in urban subscribers. Going forward, with a shift in focus towards rural areas, it is logical to believe that rural subscriber growth will continue to outpace urban subscriber growth.

Although rural areas would be the next growth driver, operators are expected to face certain challenges:

- **High network OPEX:** Operation and maintenance costs of cell sites in rural areas are high due to lower availability of electricity and thus operators are forced to depend on diesel for power supply. According to the FICCI-BDA wireless broadband report, cell site operating costs in rural areas are estimated to be 25% higher at $1,410/month vs. $1,050/month in urban areas.
- **Low ARPUs and MOU:** Due to lower per capita income and low usage, ARPUs and MOU are lower in rural areas than in urban areas.

Apart from huge CAPEX and OPEX requirements, other constraints in increasing penetration in rural areas include acquisition of land, unavailability of cheap and fast backhaul connectivity, lack of continuous power supply, and low literacy levels. To increase rural penetration, TRAI has taken numerous initiatives such as recommending bringing mobile services under the ambit of USOF and sharing of infrastructure to receive support from USOF, and supporting backbone infrastructure through USOF. Therefore, low penetration, coupled with factors such as increasing affordability, lower handset prices and TRAI initiatives, suggests significant potential in rural areas. It is possible that the operators will expand coverage in rural areas to gain the first-mover advantage. This in turn should help improve rural teledensity.

**Launch of new services:** Services like IPTV and DTH services are being provided by some operators. For example, to leverage its existing fixed line infrastructure, Bharti Airtel has started to offer IPTV and DTH services. While IPTV is limited initially to few cities, the company believes that DTH would become a mass product. Bharti will provide bundled services, commonly known as *triple play* – voice, broadband and TV services on one platform.

**Declining ARPUs to put pressure on margin:** The telecom industry is seeing adds of over 10 million subscribers every month, but the increasing base is not resulting in an increase in usage, as reflected by MOU. On the revenue front, growth is slowing down with a higher base and due to the consistent fall in ARPU. Going forward, a shift in the operators' focus towards low-usage rural subscribers would further contribute to a fall in ARPU. Furthermore, the competitive pressure will likely heighten with new operators rolling out services, which in turn would lead to downward pressure on tariffs. TRAI recently reduced the termination charges for all domestic calls to INR0.2/min from INR0.3/min from April onwards, which should further aid the tariff decline, pressurizing ARPUs. This, along with slow subscriber base growth, might result in slowing top-line growth, which, together with increasing competition and higher CAPEX to expand business, should put pressure on margins. However, to scale down the effect, the wireless operators are looking to increase value added service revenue similar to the developed markets, where operators have shifted focus towards data revenue, which has led to an improvement in margins.

**Higher spectrum charges:** The GSM operators were given an initial allotment of 4.4 MHz for network roll-outs, which was bundled along with the telecom license. However, the regulator has proposed a 1% increase in usage charges for spectrum upto 8 MHz and a 2% increase for spectrum up to and above 8 MHz. While a final decision is pending, this will likely impact majority of the players, as most of them have spectrum over 6.2 MHz in most of the Circles.

**Future growth expected in non-voice (data) revenue in highly penetrated areas:** Circle-wise, penetration levels are high in metros and A Circles in terms of voice telephony. These Circles are associated with high ARPUs. Therefore, with the voice telephony market nearing maturity in these areas, ARPUs are declining at a much faster pace. As ARPUs decline and voice becomes commoditized, the challenge for the telecom industry would be to retain customers, develop alternative revenue streams, and create a basis for differentiation in high-churn markets. This would in turn result in mobile operators shifting towards non-voice revenue, similar to the developed markets, where operators have shifted towards data revenue. It seems that data revenue will continue to offset the loss from voice ARPU. Moreover, it is rising as a percentage of wireless revenue. Globally, mobile data revenue growth is increasing and the voice market has been seeing a decline. In India as well, a similar trend is being observed, especially in the metros and urban areas, where mobile operators have shifted focus from customer acquisition to promoting value-added services. They are also looking at data as the next growth driver and a significant source of revenue. However, at present, data revenue contributes just 8- 10% to the operators' overall revenue. Data revenue growth is led by SMS, with ring tones and caller ring-back tones being the next largest driver. Furthermore, 3G, which is expected to be the driving factor for high-end data services, should also help boost data revenue in India.

**Infrastructure sharing:** Recognizing the critical importance of infrastructure to offer wireless telecom services, TRAI has been taking various initiatives to encourage infrastructure sharing. TRAI considered the passive infrastructure sharing issue and recommended active infrastructure sharing and backhaul in April 2007. Passive infrastructure constitutes about 60% of the total cell site cost. Therefore, sharing would enable operators to reduce CAPEX to a large extent. Passive infrastructure sharing has provided a new business stream for telecom players and they appear to be realizing the benefits. Large players with a pan-India footprint have hived off their tower infrastructure units into companies to unlock value. For new players, who are expanding network, passive infrastructure sharing entails lower CAPEX and faster roll-out of services.

According to TRAI's estimates, India would need about 350,000 towers by 2010 to cater to a target subscriber base of 500 million. Due to strong demand for towers, given a rising subscriber base, tower companies have set aggressive target roll-outs. Many independent tower companies have also announced aggressive tower roll-outs to capture the growth opportunity. By 2010, Bharti Infratel plans to have 115,000 towers and Reliance plans to follow suit with 63,000 towers, GTL Infrastructure is also planning 25,000 towers. This is about 50% of the expected 350,000 towers.

Such aggressive tower roll-outs will likely result in an oversupply over a short period of time. However, the existing operators' expansion plans and the entry of new players should create demand for towers. It is likely that most of the new players will lease towers to ensure faster roll-out of networks rather than incur CAPEX. Further, with large number of independent tower companies and integrated telecom players in the industry, one can only expect consolidation in the tower industry, signs of which are already visible.

## 10.3 Challenges

Some of the challenges for telecommunication industry in India are: regulatory uncertainties, severe competition, lack of affordability for the rural consumers, infrastructure and social issues. These challenges are briefly discussed below. The two biggest external challenges for telecommunication industry in India are regulatory uncertainty and competition.

**Regulatory uncertainty:** Lack of a clear regulatory roadmap has been one of the major challenges for the Indian telecom industry. Uncertainty about the policy and regulatory framework revolve around the following:

- **Final decision on spectrum not yet taken:** With TRAI and TEC increasing the subscriber base criteria, it has now become difficult for the existing operators to get additional spectrum. For efficient and quality services, spectrum availability is a must. As these norms for spectrum allocation are implemented, the existing operators will have to increase their subscriber base to get additional spectrum. However, a final decision hasn't been taken yet. The scarcity of spectrum is an industry-wide issue, which needs to be addressed. It may affect smaller players more because huge CAPEX is required to upgrade the network due to the spectrum crunch. The extent to which an operator would be affected would depend on its capability to maintain the quality of its services, which becomes even more important in the wake of the introduction of mobile number portability (MNP). From a subscriber perspective, it seems that net additions will not be affected much with the spectrum crunch. The criteria for subscribing a mobile connection do not include determining the amount of spectrum with a service provider. However, the existing subscribers may switch over to other providers due to deterioration in QoS. With the launch of MNP, switching over to another service provider would be easier. Another impact would be increased competition between service providers. This may result in aggressive pricing strategies to retain subscribers. Spectrum scarcity is a fundamental problem. Not much of spectrum is available for commercial cellular services despite most service operators having already become eligible for additional spectrum. Moreover, the Defense Ministry has delayed vacating of spectrum for a long time now.

- **Delay in 3G auctions:** India has been waiting for the launch of 3G services for a long time now. Although the government has announced a comprehensive policy for 3G, it has been surrounded by controversies, leading to a series of amendments. Procedural delays and disagreements between the Finance Ministry, DoT, and TRAI on the reserve price for spectrum auctions have been delaying the schedule. The government has not been able to lay down a clear roadmap for 3G auctions. Furthermore, the delay in auctions is benefiting state-owned companies such as BSNL and MTNL, as spectrum has been allocated to them, whereas private players will have to wait for the auctions to begin.

- **Uncertainty about defense vacating the spectrum:** The signing of an MOU between the DoT and the Defense Ministry for vacating spectrum has been delayed. The Defense Ministry earlier agreed to release 10 MHz of spectrum for third generation mobile services and another 5 MHz for existing GSM operators immediately after signing the agreement. The DoT, on its part, agreed to lay an optical fiber cable for the armed forces, connecting over 270 locations over three years. However, with the Defense ministry now proposing new conditions in the MoU, the vacation of spectrum has been delayed. These conditions include powers to take back the vacated spectrum if DoT fails to roll out an alternative optical fiber cable network for the armed forces within the stipulated period. It has also proposed that spectrum should be reverted to the armed forces in case operators do not start using it within a specified time. The DoT has so far not agreed to these conditions, which creates uncertainty about spectrum availability for mobile operators. Thus, an increasing delay in signing of the MoU could put off the 3G auctions further and also increase congestion in existing cellular networks.

- **Uncertainty about introduction of MNP:** In August 2008, DoT announced the guidelines for implementation of intra-circle MNP with a timeline of mid-09 for metros and end-09 for the whole country. The DoT divided the country into two geographic zones, each of which would be handled by a different MNP provider. Each zone was further broken down into 11 service areas that represent cities within the zone. Zone 1 will cover the northern and western regions, while Zone 2 will include the south and east. The centralized operators for implementing MNP were finalized in March. Syniverse will cover northern and western states, while MNP Interconnection Telecom Solutions India, a joint venture of Telcordia, will cover the southern and eastern states. However, these centralized operators are required to roll out MNP in metros and A-category Circles within six months of receiving the license, and extend it to the other circles in the next six months. Therefore, there is uncertainty about the implementation of MNP across the country by December 2009.

**Competition:** To take advantage of the boom in the telecom sector, many new aspirants applied for licenses in the past. The DOT received as many as 575 applications. However, letters of intent were given to only those who applied before the due date. Apart from leading global telecom giants, those with presence in real estate and infrastructure sectors applied for licenses to be a part of the growth story. The DOT issued letters of intent to nine different companies. Of them, only two-Unitech and STel-were completely new. Others were Tata Teleservices, Loop Telecom (owned by BPL), Spice Communication, Shyam Telelink, Swan Telecom, Datacom and Idea Cellular. These letters of intent have now been converted into licenses. With the issue of start-up spectrum allocated to the new licensees, the number of operators in a particular Circle would be more than four. However, the entry of new players would not pose any threat to existing operators, as the chances of them affecting the market in the near term are low due to significant market and execution challenges. Apart from spectrum constraints, these players would face financing risks and severe competition, as they are late entrants and are likely to enter simultaneously in an already crowded market.

**Lack of affordability:** When considering any technology for rural India, the issue of affordability must be addressed first. Given the income level, one must determine the cost of any sustainable solution. It is reasonable to expect an expenditure on telecommunication services of only around USD 1.5 per month on the average (2 per cent of household income). In this scenario, mobile handsets with advanced features will be too expensive and beyond afford for majority of the rural population.

**Lack of infrastructure:** In many villages in India availability of uninterrupted supply of electricity is not there. This causes a hindrance to the spread on Internet in rural India. Moreover, rural India suffers from poor availability of backhaul connection. Finally, the last mile in rural India does not exist. The cost of laying down last mile is prohibitively expensive.

**Social issues:** For many rural families in India, phone is not a personal object. Mobile phones, like a land line, are used by the entire family. Unless this attitude of rural family changes, the rural teledensity will remain low.

# 11 Research Challenges

The design of next generation converged networks will pose a number of significant research challenges. There are several crucial issues that need to be investigated before a true convergence can happen in the network and service perspective. Some of the important issues are enumerated below:

- **Architectural Framework:** The migration towards NGN changes the architecture and topology of networks which potentially involves several structural changes, such as a reorganization of core network nodes and changes in the number of network hierarchy levels. The shift to IP networks also raises questions whether interconnection frameworks need to be revised.

- **QoS issues:** NGN will be all-IP based and will have to provide guaranteed QoS to mobile terminals. QoS provisioning in a heterogeneous wireless and mobile networks will bring in new problems to mobility management, such as location management for efficient access and timely service delivery, QoS negotiation during inter-system handoff, etc.

- **Design of user terminals:** The design of single user terminal that is able to autonomously operate in different heterogeneous access networks will be another important research challenge. This terminal will have to exploit various surrounding information (e.g., communication with localization systems, cross-layer with network entities etc.) in order to provide richer user services (e.g., location/situation/context-aware multimedia services). This will also put strong emphasis on the concept of cognitive radio and cognitive algorithms for terminal re-configurability.

- **Location and handoff management in wireless overlay networks:** Future wireless networks will be inherently hierarchical where access networks have different coverage areas. Mobility management in wireless overlay networks will pose a difficult challenge to solve.

- **Cross-layer optimization:** Design of efficient cross-layer-based approaches will be instrumental in developing new mobility management schemes. It has already been observed through research that cooperation between the network and link layers is able to improve the performance of mobility management in IP-based heterogeneous communication environment. Information from the link layer such as signal strength and velocity of mobile terminals may help the decision making of mobility management techniques at the network layer. In cross-layer optimization, how to cooperate, how tight the cooperation is, and how much information is to be exchanged between the two layers are possible research issues.

- **Other issues:** Efficient use of spectrum, Fault-tolerance, availability of network services, enhanced security, intelligent packet and call routing, intelligent gateway discovery and selection protocol design and development of a unified protocol stack and vertical protocol integration mechanisms are some of the other important research issues in next-generation heterogeneous networks.

## 12 Conclusion

This chapter has provided a clear picture of a converged, all IP communications environment, which fulfils almost all the expectations and requirements of a NGN system. The aim was to help the reader comprehend the concept of convergence, its drivers and enablers in NGN. Various issues of NGN are discussed and the current and future trends of standardization activities for NGN are presented in detail. IMS is depicted as a major enabler for achieving convergence. The paper had presented a case study to illustrate how IMS can be used to achieve convergence in a residential networking environment. Some of the applications mentioned may be a bit advanced and not feasible from the perspective of India, they are certainly going to be deployed in near future. The emerging trends of the Indian telecommunication industry are also discussed and some of the challenges that the industry is facing today for its growth and evolution are presented. Finally some open problems in the domain of NGN and convergence are discussed for the potential researchers to explore.

# Appendix A - List of Abbreviations

| | |
|---|---|
| 3GPP | Third Generation Partnership Project |
| 3GPP2 | Third Generation Partnership Project 2 |
| AAA | Authentication Authorization and Accounting |
| ADC | Access Deficit Charge |
| ADSL | Asymmetric Digital Subscriber Line |
| AOL | America on Line |
| API | Application Programming Interface |
| ARPU | Average Revenue per User |
| AS | Application Server |
| ASP | Application Service Provider |
| ATIS | Alliance for Telecommunication Industry Solutions |
| B2BUA | Back to Back User Agent |
| BBC | British Broadcasting Corporation |
| BDA | Broadband Development Authority |
| BGCF | Breakout Gateway Control Function |
| BPL | Broadband over Power Line |
| BSNL | Bharat Sanchar Nigam Limited |
| BWA | Broadband Wireless Access |
| CAGR | Compound Annual Growth Rate |
| CAMEL | Customized Applications for Mobile network Enhanced Logic |
| CAP | CAMEL Application Part |
| CAPEX | Capital Expense |
| CATV | Cable Television |
| CDMA | Code Division Multiple Access |
| CDR | Charging Data Record |
| CN | Core Network |
| COPS | Common Open Policy Service |
| COPS- PR | Common Open Policy Service- Policy Provisioning |
| CPNP | Calling Party's Network Pays |
| CS | Circuit Switch |
| CSCF | Call Session Control Function |
| DiffServ | Differentiated Service |
| DNS | Domain Name System |
| DOCSIS | Data over Cable Service Interface Specifications |
| DoS | Denial of Service |
| DoT | Department of Telecommunications |
| DRM | Digital Rights Management |
| DSCP | Differentiated Service Code Points |
| DSL | Digital Subscriber Line |
| DSLAM | Digital Subscriber Line Access Multiplexer |
| DSL-F | DSL Forum |
| DTH | Direct to Home |
| DTMF | Dual Tone Multi Frequency |
| DTT | Digital Terrestrial Television |
| ESIF | Emergency Services Interconnection Forum |
| ESP | Encapsulated Security Payload |
| ETS | Emergency Telecom Service |
| ETSI | European Telecommunications Standards Institute |
| EU | European Union |
| FCC | Federal Communications Commission |
| FICCI | Federation of Indian Chamber of Commerce and Industry |
| FMC | Fixed–Mobile Convergence |
| GERAN | GSM EDGE Radio Access Network |
| GPRS | General Packet Radio Service |
| GSM | Global System for Mobile communications |
| HDTV | High Definition Television |
| HFC | Hybrid Fiber Copper |
| HSS | Home Subscriber Server |
| HTTP | Hyper Text Transfer Protocol |
| I-CSCF | Interrogating- Call Session Control Function |
| IEEE | Institute for Electrical and Electronics Engineer |
| IESG | Internet Engineering Steering Group |
| IETF | Internet Engineering Task Force |
| IKE | Internet Key Exchange |
| IM | Instant Messaging |
| IMS | IP Multimedia Subsystem |
| IMS-ALG | IMS Application Level Gateway |
| IM-SSF | IMS Switching Function |
| INC | Industry Numbering Committee |
| IP | Internet Protocol |
| IP-CAN | Internet Protocol- Connectivity Access Network |
| IPSec | Internet Protocol Security |
| IPTV | Internet Protocol Television |
| IPv4 | Internet Protocol version 4 |
| IPv6 | Internet Protocol version 6 |
| ISC | IMS Service Control |
| ISDN | Integrated Services Digital Network |
| ISIM | IMS Subscriber Identity Module |
| ISO | International Organization for Standardization |
| ISP | Internet Service Provider |
| ISUP | ISDN User Part |
| ITU | International Telecommunication Union |
| ITU-T | International Telecommunication Union-Telecom standardization sector |
| IVR | Interactive Voice Response |
| LAES | Lawfully Authorized Electronic Surveillance |
| LAN | Local Area Network |
| LEA | Law Enforcement Agency |
| LI | Lawful Interception |
| LMDS | Local Multipoint Distribution Service |
| LRIC | Long Run Incremental Cost |
| LTE | Long Term Evolution |
| MAN | Metropolitan Area Network |
| MBMS | Multimedia Broadcast Multicast Service |
| MEF | Metro Ethernet Forum |
| MGCF | Media Gateway Control Function |
| MGW | Media Gateway |
| MMD | Multimedia Domain |
| MNP | Mobile Number Portability |
| MOU | Memorandum of Understanding |
| MOU | Minutes of Use |
| MPLS | Multi Protocol Label Switching |
| MRF | Media Resource Function |
| MRFC | Media Resource Function Controller |
| MRFP | Media Resource Function Processor |
| MSF | Multi-service Switching Forum |
| MTNL | Mahanagar Telephone Nigam Limited |
| NAT-PT | Network Address Translation – Protocol Translation |
| NBC | National Broadcasting Company |
| NENA | National Emergency Numbering Association |
| NGN | Next Generation Network |
| NGN- FG | Next-Generation Network – Focus Group |
| NGSP | Next Generation Service Provider |
| NIC | Network Interface Card |
| NRIC | Network Reliability and Interoperability Council |
| OAMP | Operation Administration Maintenance and Provisioning |
| OMA | Open Mobile Alliance |
| OPEX | Operational Expense |
| OSA-SCS | Open Source Access- Service Capability Server |
| P2P | Peer-to- Peer |
| PCG | Project Coordination Group |
| P-CSCF | Proxy- Call Session Control Function |
| PDA | Personal Digital Assistant |
| PDF | Policy Decision Function |
| PDP | Packet Data Protocol |
| PEP | Policy Enforcement Point |
| PHB | Per Hop Behavior |
| PIB | Policy Information Base |
| PLMN | Public Land Mobile Network |
| PoC | Push to talk over Cellular |
| PRI | Provisioning Instances |
| PSTN | Public Switched Telephony Network |
| PVR | Personal Video Recorder |
| QoS | Quality of Service |
| RFC | Request for Comments |
| RSVP | Resource Reservation Protocol |
| S-CSCF | Serving-Call Session Control Function |
| SCTP | Stream Control Transmission Protocol |
| SDO | Standards Development Organization |
| SDP | Session Description Protocol |
| SEG | Security Gateway |
| SGW | Signaling Gateway |
| SIP | Session Initiation Protocol |
| SLF | Subscriber Location Function |
| SMTP | Simple Mail Transfer Protocol |
| TCP | Transmission Control Protocol |
| TDM | Time Division Multiplexing |
| TEC | Telecommunication Engineering Center |
| THIG | Topology Hiding Inter-network Gateway |
| TIA | Telecommunications Industry Association |
| TISN | Trusted Information Sharing Network |
| TISPAN | Technical committee within ETSI for Next Generation Networks |
| TR | Technical Report |
| TSG | Technical Specification Group |
| TRAI | Telecom Regulatory Authority of India |
| UAS | Universal Access Service |
| UE | User Equipment |
| UICC | Universal Integrated Circuit Card |
| UMA | Unlicensed Mobile Access |
| UMTS | Universal Mobile Telecommunication Services |
| UNI | User Network Interface |
| URI | Uniform Resource Identifier |
| USD | Universal Service Directive |
| USO | Universal Service Obligation |
| USOF | Universal Service Obligation Fund |
| UTRAN | UMTS Terrestrial Radio Access Network |
| VoD | Video on Demand |
| VoIP | Voice over Internet Protocol |
| VoWi-Fi | Wireless Fidelity enabled Voice over Internet Protocol |
| VSAT | Very Small Aperture Terminal |
| WCDMA | Wideband Code Division Multiple Access |
| WiFi | Wireless Fidelity |
| WiMAX | Worldwide Interoperability for Microwave Access |
| WLAN | Wireless Local Area Network |